\documentclass[apj]{emulateapj}
\usepackage{amsmath}
\usepackage{epstopdf}
\usepackage{dpfloat}

\usepackage{bm}
\usepackage{dpfloat}

\shorttitle{}
\shortauthors{Speights, Westpfahl}

\begin{document}

\title{The Pattern Speeds of NGC 3031, NGC 2366, and DDO 154 as Functions of Radius}

\author{Jason C. Speights, David J. Westpfahl}
\affil{Physics Department \\ New Mexico Institute of Mining and Technology\\ Socorro, NM 87801}

\begin{abstract} 

The pattern speeds of NGC 3031, NGC 2366, and DDO 154 are measured using a solution of the Tremaine-Weinberg equations derived in a previous paper.  Four different data sets of NGC 3031 produce consistent results despite differences in angular resolution, spectral resolution, and sensitivities to structures on different scales.  The results for NGC 3031 show that the pattern speed is more similar to the material speed than it is to the speed of a rigidly rotating pattern, and that there are no clear indications of unique corotation or Lindblad resonances.  Unlike NGC 3031, the results for NGC 2366 and DDO 154 show clear departures from the material speed.  The results for NGC 2366 and DDO 154 also show that the solution method can produce meaningful results that are simple to interpret even if there is not a coherent or well-defined pattern in the data.  The angular resolution of a data set has the greatest affect on the results, especially for determining the radial behavior of the pattern speed, and whether there is a single, global pattern speed.

\end{abstract}


\section{INTRODUCTION}

The purpose of this paper is to show how different properties of a data set affect the results from using a method developed in a previous paper (Speights \& Westpfahl 2011, hereafter Paper I) for solving the pattern speed ($\Omega_p$) equations of Tremaine \& Weinberg (1984, hereafter TW84).  This is important for choosing data in future applications of the solution method, with the long-term goal of determining whether spiral patterns are rigidly rotating waves, shearing material arms, or perhaps some combination of both. 

The pattern speed is traditionally determined from a guided interpretation of observations according to the density wave theory (Lin \& Shu 1964, 1966; Lin et al. 1969).  In density wave theory $\Omega_p$ is a free parameter that is measurable by assuming that it is constant (i.e., independent of radius), and that resonant radii are identifiable from photometric and kinematic features of the galaxy.  Resonance is assumed to occur when $\Omega_p$ = $\Omega$, the material speed (corotation resonance), and when $\Omega_p$ = $\Omega$ $\pm$ $\kappa$/$m$, where $\kappa$ is the epicycle frequency of a stellar orbit and $m$ is the mode of the wave corresponding to the number of arms (Lindblad resonance).  The assumption that $\Omega_p$ is constant is appealing if one assumes that spiral patterns persist for many dynamical (orbital) timescales because this avoids the winding dilemma for material arms.

The TW84 equations, however, are independent of the detailed assumptions of spiral structure theories, and solutions can be used to test those assumptions.  These are integrated forms of the continuity equation that relate the pattern speed to the observable kinematic and density properties of a pattern tracer.  The derivation in TW84 assumes that the disk is flat, that the pattern speed is well defined, and that the pattern tracer obeys the continuity equation.  

In Paper I the TW84 equations are solved for a pattern speed that is allowed to vary with radius.  The solution method is then applied to the H{\hskip 1.5pt \footnotesize  I} spiral pattern of the barred, grand-design galaxy NGC 1365.  A statistical analysis of the results rules out the hypothesis of a constant $\Omega_p$ for the spiral arms.  The pattern speed behaves approximately as 1/$r$, and is very similar to the material speed.  There are also no clear indications of unique corotation or Lindblad resonances for the spiral arms.  It is noted in Paper I that the winding dilemma is avoidable for NGC 1365 if the spiral pattern is a transient feature.

The results for NGC 1365 are inconsistent with traditional methods for measuring spiral arm pattern speeds.  More direct measurements that allow for a radially varying $\Omega_p$ are therefore needed in order to estimate how common shearing patterns are.  There is an abundant amount of archived data for doing so, and a better understanding of how different properties of a data set affect the results will help with selecting useful data.  

This paper reports the results from applying the solution method of Paper I to the grand-design galaxy NGC 3031, the irregular galaxy NGC 2366, and the dwarf irregular galaxy DDO 154.  Four different data sets of NGC 3031 show how angular resolution, spectral resolution, and sensitivities to structures on different scales affect the results.  The pattern speed is found for NGC 2366 and DDO 154 to show how the results are affected in the absence of a coherent or well-organized pattern.  The results for the irregular galaxies also provide a counterexample to the results for NGC 3031. 

The rest of this paper is organized as follows.  In Section 2 the solution method developed in Paper I is briefly reviewed, and a new procedure is introduced for selecting models that are an alternative to a constant $\Omega_p$.  In Section 3 the method of analyzing the results is explained.  In Section 4 the data are described.  In Section 5 the results are presented and analyzed.  In Section 6 the results are discussed.  In Section 7 is a summary.

\section{SOLUTION METHOD}

When $\Omega_p$ is allowed to vary with radius in the TW84 equations one of the results is an integral equation,
\begin{eqnarray}
\nonumber \int_{-\infty}^{+\infty}{\Omega_p(r)\Bigl (x{\partial \over \partial y}I(x,y) - y{\partial  \over \partial x}I(x,y)\Bigr)}&dx \\  =  \int_{-\infty}^{+\infty}{\partial \over \partial y}I(x,y) \,v_y(x,y)\, & dx,
\end{eqnarray}
that is solvable for simple functional forms of $\Omega_p$.  In Equation (1), $I$ is the intensity of a pattern tracer and $v_y$ is the line of sight velocity of the tracer (corrected for the systemic velocity, $V_{\mbox{\scriptsize sys}}$) times the cosecant of the disk inclination, $i$.  The $x$ and $y$ coordinates are measured relative to the kinematic minor and major axis respectively.   For an illustration showing how a galaxy disk is oriented in the coordinate system of Equation (1), see Figure (1) in Paper I.  

The two forms of $\Omega_p$ that are used in Paper I for solving Equation (1) are a polynomial in $r$ = $\sqrt{x^2+y^2}$,
\begin{equation}
\Omega_p(r) \, =  \sum_{i=0}^{n}\alpha_i r^{i},
\end{equation}
and a polynomial in 1/$r$,
\begin{equation}
\Omega_p(r) \, =  \sum_{i=0}^{n}{\alpha_i \over r^{i}},
\end{equation} 
where that $n$ is an integer and $\alpha_i$ are unknown coefficients that are determined in a solution.  Both Equations (2) and (3) can represent a constant $\Omega_p$ when $n$ = 0, and higher order terms can be used to determine whether the pattern is rigid or shearing.  Determining the order at which point it is appropriate to truncate the summations in Equations (2) and (3) is the primary goal of the analysis in Paper I.  

The advantage of using such forms for $\Omega_p$ is that the unknown coefficients can come out of the integral on the left hand side of Equation (1).  Using Equation (2),  Equation (1) becomes, 
\begin{eqnarray}
 \nonumber \sum_{i=0}^{n}\alpha_i\int_{-\infty}^{+\infty}r^{i}\,\Big\{x{\partial \over \partial y}I(x,y) - y{\partial \over \partial x}I(x,y)\Big\}&dx\\ =  \int_{-\infty}^{+\infty}{\partial\over \partial y}I(x,y)v_y(x,y)\,&dx,
\end{eqnarray}
or similarly using Equation (3),
\begin{eqnarray}
 \nonumber \sum_{i=0}^{n}\alpha_i\int_{-\infty}^{+\infty}{1 \over r^{i}}\,\Big\{x{\partial \over \partial y}I(x,y) - y{\partial \over \partial x}I(x,y)\Big\}&dx \\  =  \int_{-\infty}^{+\infty}{\partial\over \partial y}I(x,y)v_y(x,y)&dx.
\end{eqnarray}
Applying either Equation (4) or (5) independently (i.e., at least a beam size apart in the case of radio synthesis maps) to different parts of the galaxy transforms the integral equation into a matrix equation of the form $\bf{G}\boldsymbol \alpha$ = $\bf{d}$.  That the calculations are independent is important because this is assumed in the methods used for estimating and propagating uncertainties, and in the statistics used for analyzing the results.  The matrix equation is easily solved using the normal equations, ${\bf{G}}^{T}\bf{G}\boldsymbol \alpha$ = ${\bf{G}}^{T}\bf{d}$, for the linear least-squares problem.  Methods for the analysis of the results for linear least-squares problems are well established, and this is another advantage of using such forms for $\Omega_p$.

The results for NGC 1365 demonstrate that only a few terms are needed in the above polynomials in order to determine the radial behavior of $\Omega_p$.  Including many terms can unnecessarily complicate the analysis, and by $|n|$ = 3 the matrix ${\bf{G}}^{T}\bf{G}$ typically becomes ill-conditioned for numerical inversion.  When ${\bf{G}}^{T}\bf{G}$ is ill-conditioned, the solution is unstable, and any inferences based on such solutions may be unreliable.  By summing Equations (2) or (3) to order $n$ = 3 one can infer whether there is any sign of a discontinuity in $\Omega_p$ such as a step function.  This is useful in Paper I because the bar in NGC 1365 may have a separate $\Omega_p$ than that of the spiral arms (Sellwood \& Sparke 1988; Sellwood 1993 and references therein), and it needed to be shown whether the pattern speed of the bar is detected in the results.  

Since the pattern of NGC 3031 shows no obvious discontinuities or signs of a bar (see Figures 1 through 4), and in order to simplify the analysis for all three galaxies, only two terms are used in this paper as an alternative to a constant $\Omega_p$ model.  The alternative models have the form,  
\begin{equation}
\Omega_p(r) = \alpha_0 + \alpha_1 r^{n},
\end{equation}
but without the restriction to integer values of $n$.  Relaxing the restriction to integer values of $n$ allows for different amounts of curvature in $\Omega_p$.  The value of $n$ is determined by comparing the goodness of fit for solutions of Equation (1) using Equation (6) and a range of different values for $n$.  For each solution, the distribution of the residuals, normalized by the propagated uncertainties, is compared with a standard normal distribution.  If one assumes that the uncertainties are well known, the distribution of the normalized residuals will be a standard normal distribution when the true form of $\Omega_p$ is used to find a solution.  Therefore, the value of $n$ that results in a distribution of the normalized residuals that is the most similar to a standard normal distribution is used to chose an alternative model for subsequent analysis.  

A Kolmogorov - Smirnov ($KS$) test is used to quantify how similar the distribution of the normalized residuals is to a standard normal distribution.  The $KS$ test consists of finding the largest distance between two cumulative distribution functions (cdf), $D$ = Max$|$cdf($R$) $-$ cdf($SN$)$|$, where that in this case $R$ is the distribution of the normalized residuals and $SN$ is a standard normal distribution.  Then, the probability, $P_{KS}$, is found for obtaining a value of $D$ should the null hypothesis be true that $R$ is a standard normal distribution (Kendall \& Stewart 1973, Chapter 30; Lupton 1993, Chapter 19).  If $P_{KS}$ is sufficiently small, then there is convincing evidence that $R$ $\neq$ $SN$, and one can infer that the model in question is not the true model.  The following convention is adopted for interpreting values of $P_{KS}$: $P_{KS}$ $\lesssim$ $1\%$ is convincing evidence that $R$ $\neq$ $SN$; $P_{KS}$ $\sim$ $5\%$ is suggestive of $R$ $\neq$ $SN$, but inconclusive; and larger values cannot rule out the null hypothesis that $R$ is a standard normal distribution.  In this paper a lower limit of 0.01\% is chosen as a cutoff for reporting small probabilities.  As a compliment to the analysis explained in the next section, the $KS$ test is also applied to the results from using a constant $\Omega_p$ model.
   
For the purpose of determining $n$, a larger value of $P_{KS}$ is considered an indication that the distribution of the normalized residuals is more similar to a standard normal distribution.  In Section 5 it is shown that two peaks consistently occur in plots of $P_{KS}$ vs $n$, one each for negative and positive values of $n$.  The results are analyzed for each case, where the model using a negative value of $n$ = $n_1$ is referred to as alternative model 1 (AM1), and the model using a positive value of $n$ = $n_2$ is referred to as alternative model 2 (AM2).

\section{ANALYSIS}

The goal of the analysis is to determine whether the pattern is rigid or shearing.  This is based on a comparison of reduced chi-square ($\chi^{2}_\nu$ for $\nu$ degrees of freedom) for a solution using a constant $\Omega_p$ model with the values obtained for AM1 and AM2.  Unless otherwise noted in Section 5, the validity of using $\chi^{2}_\nu$ as an indicator of the goodness of fit is based on the results of a Lilliefors test for quantifying how similar $R$ is to a normal distribution.  The Lilliefors test follows the same procedure as the $KS$ test, with the difference that  $D$ = Max$|$cdf($R$) $-$ cdf($N$)$|$, where that $N$ is a normal distribution with a mean and a standard deviation that is the same as that for $R$ (Mandansky 1988, Chapter 1).  If the probability, $P_L$, of obtaining $D$ is sufficiently small, then $R$ may not have a normal distribution, and the value of $\chi^{2}_\nu$ may not be a reliable indicator of the goodness of fit.  The same convention that is adopted for interpreting values of $P_{KS}$ is adopted for interpreting values of $P_L$.  

If a solution using AM1 or AM2 has a value of $\chi^{2}_\nu$ that is an improvement to the value of $\chi^{2}_\nu$ for the constant $\Omega_p$ model, then Fisher's $F$ test of an additional term is used to determine if the improvement is statistically significant.  The $F$ test is a natural choice for quantifying the significance of the improvement in $\chi^{2}_\nu$ because the constant $\Omega_p$ model is a reduced version of AM1 and AM2.  The $F$ test consists of calculating $F$ $=$ ${\triangle\chi^{2}}$/${\chi^{2}_\nu}$ and finding the probability, $P_{F}$, for obtaining such a value of $F$ should the null hypothesis be true that the extra term, $\alpha_{1}$, is zero (Bevington \& Robinson 2002, Chapter 10).  If $P_F$ is sufficiently small, then there is convincing evidence that $\alpha_1$ $\neq$ 0, and one can infer that the pattern is shearing.  If such evidence is found then the result for the constant $\Omega_p$ model is the mean of $\Omega_p$.  The following convention is adopted for interpreting values of $P_{F}$:  $P_{F}$ $\lesssim$ $1\%$ is convincing evidence that $\alpha_1$ $\neq$ 0; $P_{F}$ $\sim$ $5\%$ is suggestive of $\alpha_1$ $\neq$ 0, but inconclusive; and larger values are an indication that the improvement is not statistically significant.     

The uncertainties for calculating $\chi^{2}_\nu$, confidence intervals for the coefficients, and confidence bands for $\Omega_p$, are estimated by propagating the uncertainties per pixel in the maps of $I$ and $v_y$, as well as the uncertainties in $x$, $y$, $i$, and $V_{\mbox{\small sys}}$ through the calculations of both {\bf G} and {\bf d}.  The method of estimating the uncertainties in the maps of $I$ and $v_y$ are adopted from the reasoning given in Paper I.  This takes into account the rms noise, the channel width, and the number of channels a typical detection occurs.  The total uncertainty for the $j$th independent calculation of Equation (1) is $\sigma^{2}_j$ $=$ $\sigma^{2}_{d_j}$ $+$ $\sum_{i=0}^{n}\sigma_{G_{j,i}}^{2}\alpha_i^{2}$ (Bevington \& Robinson 2002, Chapter 6).  The confidence intervals and confidence bands are not estimated using the covariance matrix because of the uncertainties in {\bf G}.  Instead, these are estimated using Monte Carlo methods.   

An incorrect value for the position angle of the disk (PA) can potentially cause large errors in solutions of the TW84 equations (Debattista 2003; Debattista \& Williams 2004).  The PA does not explicitly appear in Equation (1), and consequently the uncertainty in the PA is not accounted for in the calculation of the propagated uncertainties.  How an incorrect PA could affect the conclusions in this paper is investigated in Section 5 by changing the PA and applying the solution method and analysis.  This is done for a range of -2$\sigma_{\mbox{\tiny PA}}$ $\leqslant$ $\bigtriangleup$PA $\leqslant$ 2$\sigma_{\mbox{\tiny PA}}$ where that $\sigma_{\mbox{\tiny PA}}$ is the 1$\sigma$ uncertainty for the adopted PA.  In Paper I it is explained that this method is preferred to Monte Carlo methods because Monte Carlo methods typically produce non-gaussian results for the distributions of the coefficients.  

In order to minimize the uncertainties in the analysis, instrumental units of [$\Omega_p$] = km s$^{-1}$ arcsec$^{-1}$ are used for solving Equation (1) and analyzing the results.  This is preferred to units commonly used for galactic disks, [$\Omega_p$] = km s$^{-1}$ kpc$^{-1}$, because the conversion from arcseconds to kpc adds a large amount of uncertainty to $x$ and $y$ owing to the uncertainty in the distance to the galaxy.  Furthermore, units commonly used for galactic disks are not necessary for the purpose of this paper.  

\section{DATA SETS}

Radio synthesis intensity and velocity maps of H{\hskip 1.5pt \footnotesize  I} are used for applying the solution method.  For a discussion of the advantages of using H{\hskip 1.5pt \footnotesize  I} as a pattern tracer the interested reader is referred to Paper I.  Although TW84 use a form of the continuity equation that assumes mass conservation, the TW84 equations are derivable form the general form of the continuity equation that includes a source function for the creation and destruction of material (Westpfahl 1998).  If the source function is constant within dynamical timescales, then it does not affect the calculation of the pattern speed (Paper I).  For the purpose of this paper it is assumed that the source function is approximately constant within dynamical timescales.  

The disk parameters that are adopted for measuring $\Omega_p$ are provided in Table 1.  The values for the kinematic centers are adopted from Trachternach et al. (2008).  The values for $i$, PA, and $V_{\mbox{\small sys}}$ are adopted from de Blok et al. (2008) for NGC 3031, and from Oh et al. (2011) for NGC 2366 and DDO 154.

\begin{deluxetable*}{llccccc}[ht!]
\tablecaption{Adopted Disk Parameters}
\tablewidth{0pt}
\startdata
\tableline\tableline\\[-7.5 pt]
Galaxy &&\multicolumn{2}{c}{Kinematic Center}&&&\\
\cline{3-4}
&& R. A. (J2000)& Dec. (J2000)& $i$ & PA & $V_{\mbox{\scriptsize sys}}$ \\
& & ($^{h}$ $^{m}$ $^{s}$) & ($\degr$ $\arcmin$ $\arcsec$) & ($\degr$) &($\degr$) &(km s$^{-1}$) \\
 \tableline\\[-7.5 pt]
NGC 3031 && 09 55 33.5 $\pm$ 0.6 & +69 03 52.0 $\pm$ 3.9 & 59 $\pm$ 2.6 & 330 $\pm$ 1.3 & \hskip 0.3pt $-$39 $\pm$ 2.8 \\
NGC 2366 && 07 28 53.9 $\pm$ 0.7 & +69 12 37.4 $\pm$ 7.8 & 63 $\pm$ 1.6 & \hskip 4.0pt 39 $\pm$ 4.0 & \hskip 3pt 104 $\pm$ 2.0 \\
DDO 154 && 12 54 05.9 $\pm$ 0.2 & +27 09 09.9 $\pm$ 3.4 & 66 $\pm$ 2.5 & 229 $\pm$ 7.8 & \hskip 3pt 375 $\pm$ 1.5  
\enddata
\end{deluxetable*}

\begin{figure*}[ht!]
\centering
\includegraphics[width=1\textwidth]{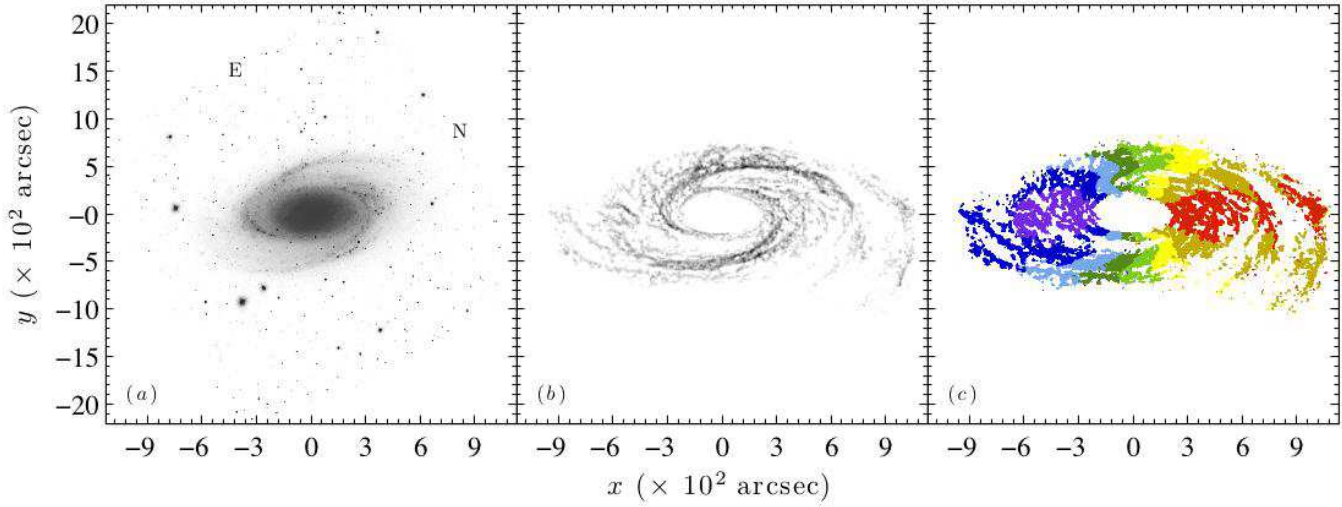} 
\caption{Data for NGC 3031 from AW96a.  Panel (a) shows is an optical image at the same scale for comparison.  Panels (b) and (c) show the maps of $I$ and $v_y$, respectively.  Directions of north (N) and east (E) on the sky are indicated in panel (a).  The receding half of the galaxy is to the right.  The peak flux is 134 Jy beam$^{-1}$ m s$^{-1}$.  The velocities are binned in increments of 67 km s$^{-1}$.} 
\end{figure*}

\begin{figure*}[ht!]
\centering
\includegraphics[width=1\textwidth]{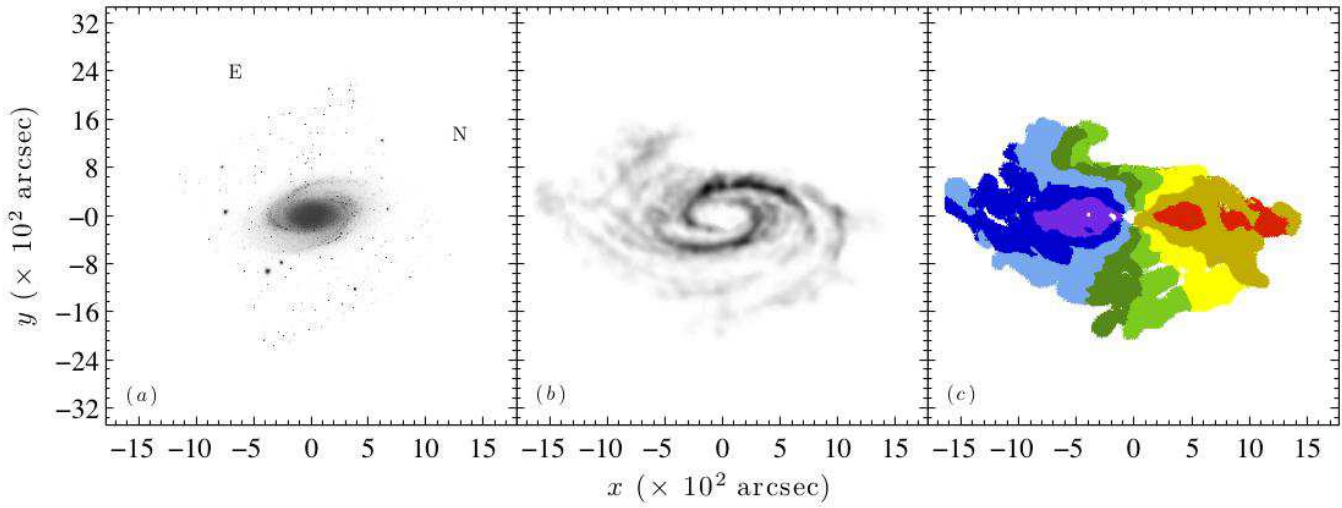} 
\caption{Data for NGC 3031 from AW96b.  The figure is formatted in the same way as Figure 1.  The peak flux is 2,880 Jy beam$^{-1}$ m s$^{-1}$.  The velocities are binned in increments of 67 km s$^{-1}$.} 
\end{figure*}

Figures 1 through 6 show the maps of $I$ and $v_y$ for the four different data sets of NGC 3031, and the data sets of NGC 2366 and DDO154, respectively.  The original maps were modified into their current form using the Astronomical Image Processing System  developed and maintained by the National Radio Astronomy Observatory\footnote{The National Radio Astronomy Observatory is operated by Associated Universities, Inc., under cooperative agreement with the National Science Foundation.} (NRAO). These are rotated and shifted from their original orientation so that their kinematic major axes are along the $x$ axis, and the origins of the $x$-$y$ coordinates are the kinematic centers that are adopted for each galaxy.  Included in the figures are optical images from the Digitized Sky Survey.  These are based on the photographic data of the National Geographic Society - Palomar Observatory Sky Survey\footnote{The NGS-POSS was funded by a grant from the National Geographic Society to the California Institute of Technology.  The plates were processed into the present compressed digital form with their permission.  The Digitized Sky Survey was produced at the Space Telescope Science Institute under US Government grant NAG W-2166.} (NGS-POSS), and are provided for comparison with the H{\hskip 1.5pt \footnotesize  I} maps.  The optical images are from the 48 inch Samuel Oschin Telescope on Palomar Mountain and a 103aE (6450 \AA) emulsion.  Table 2 provides a summary of the properties of each data set.   The data are described in greater detail in the following subsections.   

\begin{deluxetable*}{lllccc}[ht!]
\tablecaption{Properties of the Data}
\tablewidth{0pt}
\startdata
\tableline\tableline\\[-7.5 pt]
Galaxy&& Notes & Channel Seperation & Beam FWHM & rms  \\
&&& (km s$^{-1}$) & ($\arcsec$) & (mJy beam$^{-1}$) \\
 \tableline\\[-7.5 pt]
NGC 3031&& AW96a & 2.6  & 9.0 $\times$ 9.0 & \hskip -4.6pt 0.6 \\ 
NGC 3031&& AW96b & 2.6 & 60.0 $\times$ 60.0 & 1.23 \\ 
NGC 3031&& H84 & \hskip -4.5pt 10.3 & 9.0 $\times$  9.0 &  \hskip -4.6pt 2.5 \\ 
NGC 3031&& YP99 & 6.6 &135.3 $\times$ 124.8 &  \hskip -4.6pt 6.8 \\ 
NGC 2366 &&  & 2.6 & 13.1 $\times$  11.9 &  \hskip -4.6pt 0.5 \\ 
DDO 154 &&  & 2.6 & 14.1 $\times$  12.6 &  \hskip -4.6pt 0.5
\enddata
\end{deluxetable*}

\begin{figure*}[ht!]
\centering
\includegraphics[width=1\textwidth]{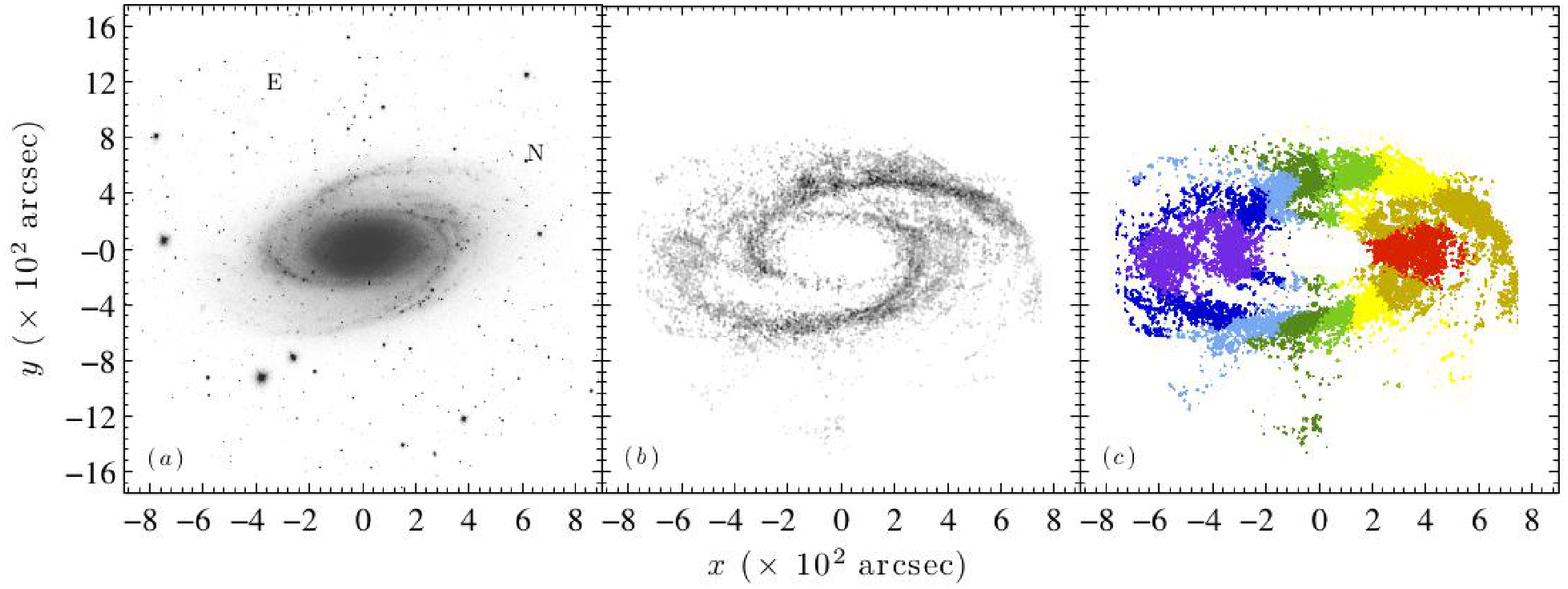} 
\caption{Data for NGC 3031 from H84.  The figure is formatted in the same way as Figure 1.  The peak flux is 497 Jy beam$^{-1}$ m s$^{-1}$.  The velocities are binned in increments of 67 km s$^{-1}$.} 
\end{figure*}

\begin{figure*}[ht!]
\centering
\includegraphics[width=1\textwidth]{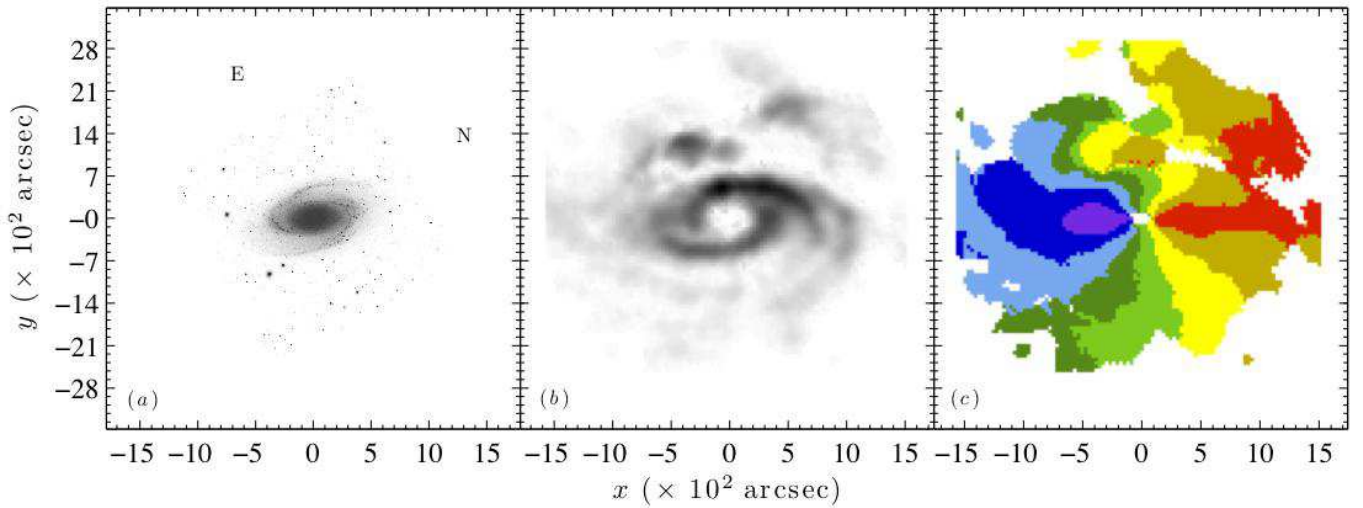} 
\caption{Data for NGC 3031 from YP99.  The figure is formatted in the same way as Figure 1.  The peak flux is 2,291 Jy beam$^{-1}$ m s$^{-1}$.  The velocities are binned in increments of 67 km s$^{-1}$.} 
\end{figure*}

\subsection{Data Sets for NGC 3031}

The four different data sets of NGC 3031 are from three different observations (Adler \& Westpfahl 1996, hereafter AW96; Hine 1984, hereafter H84; Yun \& Purton 1999, private communication, hereafter YP99).  They have different combinations of angular resolution, spectral resolution, and sensitivities to structures on different scales.  

\begin{figure*}[ht!]
\centering
\includegraphics[width=1\textwidth]{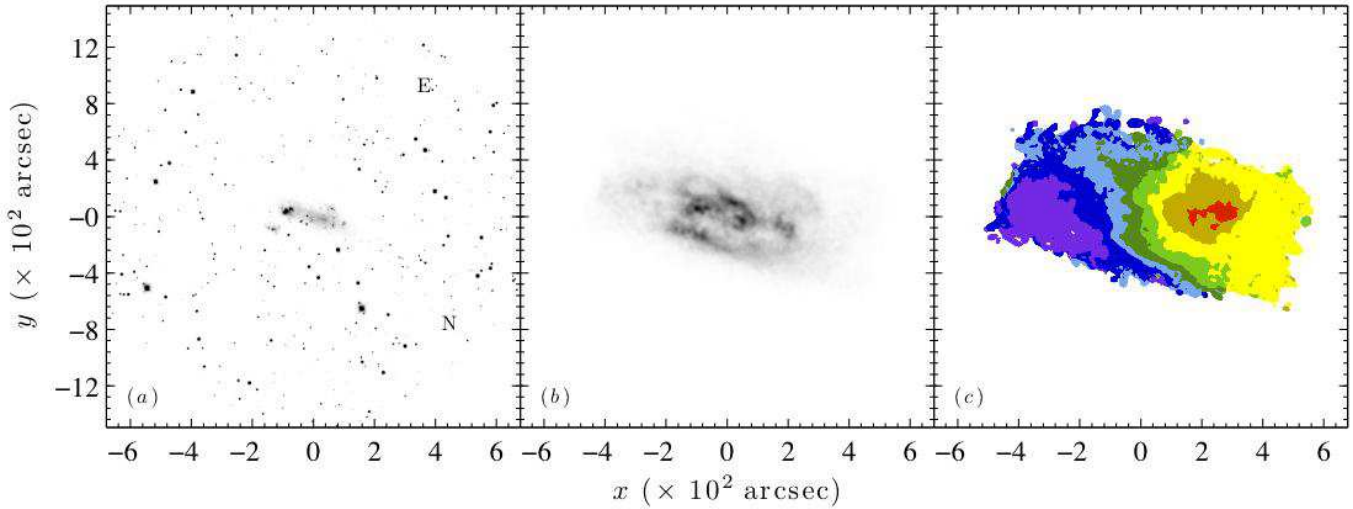} 
\caption{Data for NGC 2366.  The figure is formatted in the same way as Figure 1.  The peak flux is 1,020 Jy beam$^{-1}$ m s$^{-1}$.  The velocities are binned in increments of 15 km s$^{-1}$.  Note the non-circular velocities for the lower left quadrant of the H{\hskip 1.5pt \footnotesize  I} disk.} 
\end{figure*}

\begin{figure*}[ht!]
\centering
\includegraphics[width=1\textwidth]{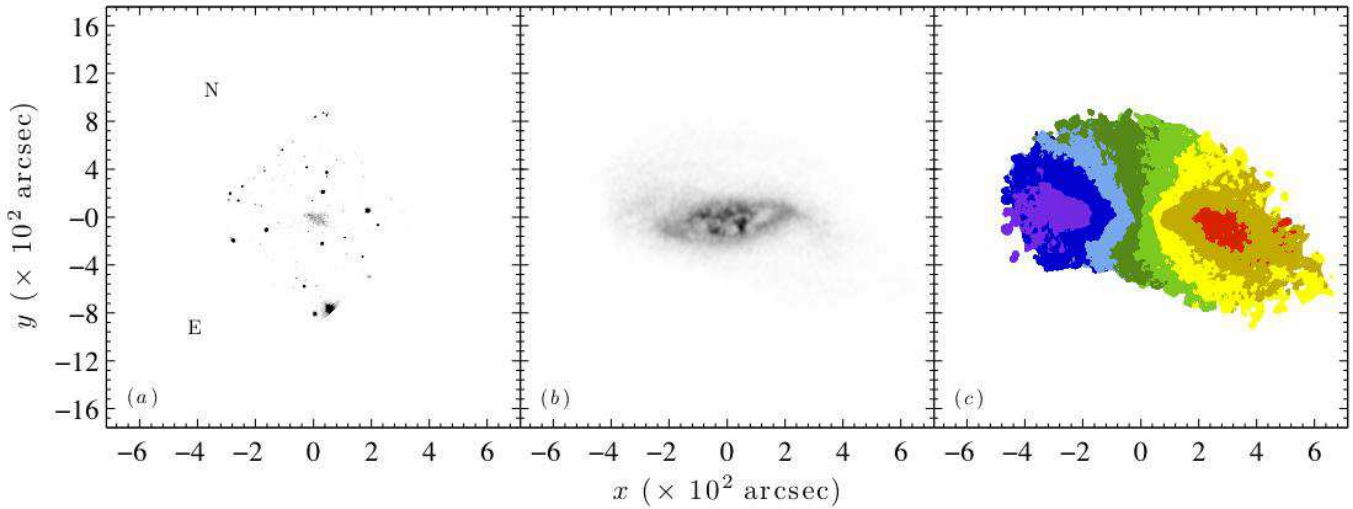} 
\caption{Data for DDO 154.  The figure is formatted in the same way as Figure 1.  The peak flux is 455 Jy beam$^{-1}$ m s$^{-1}$.  The velocities are binned in increments of 14 km s$^{-1}$.} 
\end{figure*}

Two of the four data sets, hereafter referred to as AW96a and AW96b, are made from the observations of AW96.  They used the Very Large Array (VLA) of the NRAO in 1992 and 1993 with all 27 antennas, 128 channels, one intermediate frequency (IF), and on-line Hanning smoothing.  Data from two pointings of the VLA in the B, C, and D configurations with a channel width and separation of 12.2 kHz (2.6 km s$^{-1}$) are combined to make a mosaic.  The maps published by AW96 do not have the continuum emission of the point sources subtracted.  For the AW96a data set this emission is subtracted from a calibrated and mapped data cube provided by Adler (2010, private communication).  The continuum subtracted data cube was used to make new intensity and velocity maps.  For this data set the rms signal in a line-free channel is 0.6 mJy beam$^{-1}$, and the FWHM of the synthesized beam is a 9.0$\arcsec$ circle.  The AW96b data set is from the same calibrated $u$-$v$ data, also provided by Adler (2010, private communication).  The $u$-$v$ data for this data set is tapered using a gaussian weighting function so that the synthesized beam is a 60.0$\arcsec$ circle.  For the AW96b data set the rms signal in a line-free channel is 1.23 mJy beam$^{-1}$.  The details of the observing and data reduction are described by AW96.  

The data of H84 are made from observations using the VLA in 1982.  Owing to limitations of the telescope and correlator at that time only 18 antennas and 32 spectral channels could be used with one IF.  Data from two pointings of the VLA in the B and D configurations with a channel separation of 48.8 kHz (10.3 km s$^{-1}$) are combined by H84 to make a mosaic.  The rms signal in a line-free channel is 2.5 mJy beam$^{-1}$.  The FWHM of the synthesized beam is a 9.0$\arcsec$ circle.  This data set includes the same range of baselines as does the AW96 data sets, but with relatively fewer intermediate ones due to the exclusion of the C configuration.  The details of the observing and data reduction are described by H84.  

The data of YP99 are made from observations using the synthesis telescope of the Dominion Radio Astrophysical Observatory\footnote{The Dominion Radio Astrophysical Observatory is operated as a national facility by the National Research Council of Canada.} in 1993.  The calibrated and mapped data cube is kindly made available in advance of publication.  The channel separation is 15.63 kHz (3.30 km s$^{-1}$), the channel width is 24.99 kHz (5.28 km s$^{-1}$).  Pairs of channels are averaged to improve the signal-to-noise ratio, giving final channel separations of 31.25 kHz or 6.60 km s$^{-1}$.  The rms signal in a line-free channel of this width is 6.8 mJy beam$^{-1}$.  The FWHM of the synthesized beam is 135.3\arcsec $\times$ 124.8$\arcsec$ with a beam PA of -4.4$^\circ$.  Regions of statistically significant signal in this cube are identified by first convolving each channel with a circular Gaussian of diameter 200$\arcsec$.  Regions in the convolved channel with signal of absolute value less than 2.5 mJy beam$^{-1}$ are judged to contain no significant detection, and these regions are blanked in the original, unconvolved channel.  The blanked channels are used to make the data maps.    

The different data sets emphasize different attributes of NGC 3031.  The YP99 data set excels at showing the faint outer structure, but its large synthesized beam cannot show fine detail.  The AW96a data set excels at showing fine detail, but AW96 show that it is missing flux, particularly on large scales.  The AW96b data set does not show the fine detail of the AW96a data set due to the $u$-$v$ tapering.  The H84 data set is in some ways a compromise between the AW96a and YP99 data sets.  It does not have the sensitivity to the small structures of AW96a, but it is missing little or no flux from NGC 3031 itself.  More fine detail is shown in Figure 1 for the data set from AW96a than in Figure 3 for the data set from H84 despite having the same sized synthesized beam.  This is because of the much higher sensitivity of the AW96a data set..

\subsection{Data Sets for NGC 2366 and DDO 154}

The data sets for NGC 2366 and DDO 154 are from the THINGS H{\hskip 1.5pt \footnotesize  I} survey (Walter et al. 2008).  Both data sets were made from B, C, and D observations using the VLA.  The observations of NGC 2366 are made in 2003 and 2004, and of DDO 154 in 2004 and 2005.  Both data sets have a channel width of 2.6 km s$^{-1}$ and an rms signal in a line-free channel of 0.5 mJy beam$^{-1}$.  The FWHM of the synthesized beam is 13.1$\arcsec$ $\times$ 11.9$\arcsec$ with a beam PA of -1.6$^\circ$ for NGC 2366, and 14.1$\arcsec$ $\times$ 12.6$\arcsec$ with a beam PA of -34.0$^\circ$ for DDO 154.  The details of the observing and data reduction are described by Walter et al. (2008).

\section{RESULTS}

The calculations for solving Equation (1) and analyzing the results are carried out in MATLAB\footnote{The Mathworks, Version 7.10.0.499 (R2010a), http://www.mathworks.com}.  Differentiation is performed using the Savitzky-Golay method (Savitzky \& Golay 1964).  Integration is performed by summing over a row of pixels in a map of the integrand.  The $KS$ test and the Lilliefors test are performed using MATLAB's Statistics Toolbox.  The rest of the beginning of this section explains how the results are presented.  The details of the results and their analysis are discussed in the subsections that follow.  

Plots of $P_{KS}$ vs $n$ are shown in Figure 7.  Note that two peaks consistently occur in the plots.  It is assumed that $\Omega_p$ does not have a large amount of curvature so that $n_1$ is not $\ll$ 1 and $n_2$ is not $\gg$ 1.  Out of a range of -6 $\leqslant$ $n$ $\leqslant$ 6 for which solutions are found, the peaks occur within -2 $\leqslant$ $n$ $\leqslant$ 2, so only this range is shown in the figure.  The values of $P_{KS}$ are not shown in the figure for when $n$ = 0 because the determinant of ${\bf{G}}^{T}\bf{G}$ is zero for a model of $\Omega_p$ = $\alpha_0$ $+$ $\alpha_1$.  The values of $n$, $D$, and $P_{KS}$ for the peaks in Figure 7 are summarized in Table 3.  Values of $D$ and $P_{KS}$ for the constant $\Omega_p$ model are also shown in Table 3.   The details of the results for the $KS$ test are discussed in paragraph 2 of Sections 5.1, 5.2, and 5.3.

The results for the coefficients are shown in Table 4.  The uncertainties for the coefficients are reported as the half-widths (HW) of their 95\% confidence intervals.  The coefficients are used to calculate plots of $\Omega_p$ for comparison with $\Omega$ and possible locations for Lindblad resonance.  The plots of $\Omega_p$ calculated from the coefficients are discussed in paragraph 3 of Sections 5.1, 5.2, and 5.3.

The results for the $F$ test are shown in Table 5.  An asterisk next to a value of $\chi_{\nu}^{2}$ indicates that the normalized residuals either failed a Liliefors test for normality, or that the results of the test are inconclusive.  The 1$\sigma$ uncertainties for $\chi^{2}_\nu$ are estimated from the variance, 2$\nu$, of a $\chi^{2}$ distribution.  These show how similar the values of $\chi_{\nu}^{2}$ are for a given pair of AM1 and AM2.  The details of the results for the $F$ test are discussed in paragraph 4 of Sections 5.1, 5.2, and 5.3.

\subsection{Results for NGC 3031}

Results are presented for the entire spiral pattern, which consists of the region $|y|$ $\leqslant$ 600$\arcsec$ in the H{\hskip 1.5pt \footnotesize  I} maps.  This provides 42 independent calculations of Equation (1) for the AW96a and H84 data sets, 10 independent calculations for the AW96b data set, and 4 independent calculations for the YP99 data set.  As a check, solutions are also found for three other regions: $|y|$ $\leqslant$ 300$\arcsec$, $-$600$\arcsec$ $\leqslant$ $y$ $\leqslant$ 0$\arcsec$, and 0$\arcsec$ $\leqslant$ $y$ $\leqslant$ 600$\arcsec$.  The results for the three other regions are consistent with those for the entire spiral pattern, so they are not presented separately.

In Figure 7 there are prominent peaks in the values of $P_{KS}$ for all four data sets.  The values of $P_{KS}$ are generally quite large at their peaks, with the exception of the peaks at $n_2$ for the AW96b and YP99 data sets.  The AW96a, AW96b, and H84 data sets show small $P_{KS}$ near $n$ =  0, but this is not the case for the YP99 data set.  For all four data sets the values of $P_{KS}$ for the constant $\Omega_p$ model are small enough to rule out the null hypothesis that the normalized residuals have a standard normal distribution.  This is convincing evidence that the constant $\Omega_p$ model is not the true model.  This conclusion, however, is only reliable for the AW96a, AW96b, and H84 data sets because the $KS$ test becomes unreliable when there are only a few calculations of the normalized residuals.  This may explain the discrepancy between the values of $P_{KS}$ near $n$ = 0 in Figure 7, and the small value of $P_{KS}$ for the constant $\Omega_p$ model in Table 3 for the YP99 data set.   

\begin{deluxetable*}{llcccccccccc}[b!]
\tablecaption{Results for the $KS$ Test} 
\tablewidth{0pt}
\startdata
\tableline\tableline\\[-7.5 pt]
Galaxy & Notes & \multicolumn{2}{c}{Constant $\Omega_p$}&&\multicolumn{3}{c}{AM1}&&\multicolumn{3}{c}{AM2}\\
\cline{3-4} \cline{6-8} \cline{10-12} 
 && $D$ & $P_{KS}$&& $n_1$ & $D$ & $P_{KS}$ && $n_2$ & $D$ & $P_{KS}$\\ 
 && $\times$ 10$^{-1}$ & (\%) &&& $\times$ 10$^{-1}$ & (\%) &&& $\times$ 10$^{-1}$ & (\%) \\
 \tableline\\[-7.5 pt]
NGC 3031    & AW96a & 3.64 & $<$ 0.01  && -0.89 & 1.18 & 56.4 && 0.80 & 1.46 & 30.5 \\
NGC 3031    & AW96b & 6.00 & \hskip 10.0pt  0.06  && -0.80 & 1.52 & 94.9 && 0.36 & 3.44 & 14.7 \\
NGC 3031    & H84 & 4.15 & $<$ 0.01 && -0.61 & 0.94 & 82.1 && 0.84 & 0.84 & 90.6 \\
NGC 3031    & YP99 & 7.50 & \hskip 10.0pt 0.78 && -1.42 & 2.47 & 90.9 && 0.31 & 4.99 & 18.9 \\
NGC 2366    & $\mathcal{A}$ & 3.33 &\hskip 10.0pt 0.45 && -0.11 & 1.09 & 88.4 && 0.12 & 1.01 & 93.2 \\
NGC 2366   & $\mathcal{B}$  & 6.14 & $<$ 0.01  && -0.06 & 1.58 & 82.3 && 0.08 & 1.55 & 84.2  \\
NGC 2366    & $\mathcal{C}$ & 4.02 &\hskip 10.0pt 2.09 && -0.20 &  1.38 & 93.5 && 0.25 & 1.45 & 91.2  \\
NGC 2366   & $\mathcal{D}$ & 3.32 &\hskip 10.0pt 8.89 && -0.31 & 2.09  & 55.2 && 0.41 & 1.52 & 88.0  \\
DDO 154    &  & 1.79 & \hskip 1.5pt 37.9 && -1.01 & 1.62  & 50.4 && 1.82 & 2.11 & 20.6 
\enddata
\end{deluxetable*} 

\begin{figure*}[ht!]
\centering
\includegraphics[width=1\textwidth]{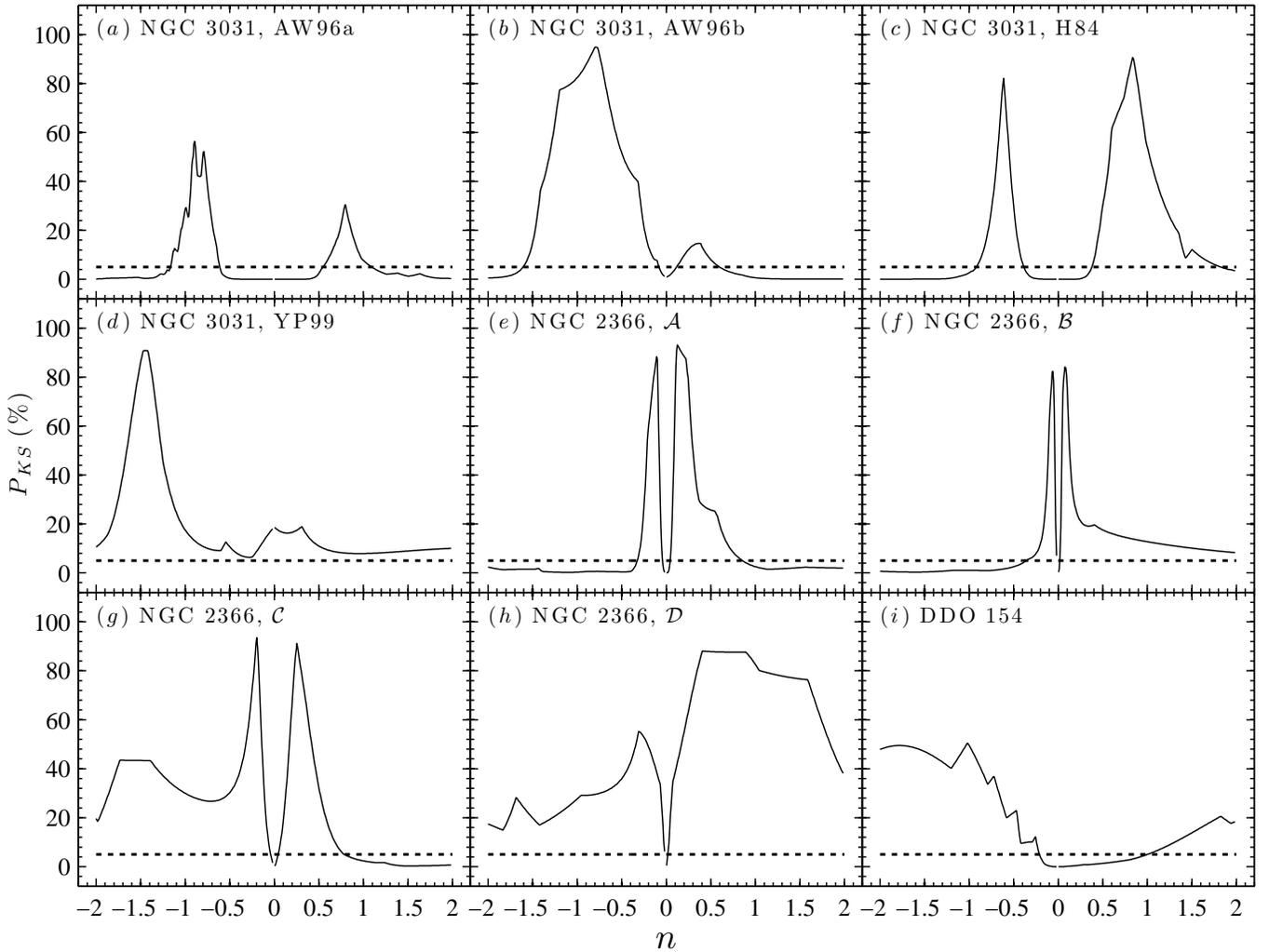} 
\caption{Plots of the results from the $KS$ test.  The solid lines are values of $P_{KS}$.  The dashed lines at $P_{KS}$ = 5\% are provided for reference.  Panels (a), (b), (c), and (d) show the results for the AW96a, AW96b, H84, and YP99 data sets of NGC 3031, respectively.  Panels (e), (f), (g), and (h) show the results for regions $\mathcal{A}$, $\mathcal{B}$, $\mathcal{C}$, and $\mathcal{D}$ of NCG 2366, respectively (See Section 5.2 for an explanation of these regions).  Panel (i) shows the results for DDO 154. } 
\end{figure*}

\begin{deluxetable*}{llccccccc}
\tablewidth{0pt}
\tablecaption{Results for the Coefficients} 
\startdata
\tableline\tableline\\[-7.5 pt]
Galaxy & Notes & Constant $\Omega_p$&&\multicolumn{2}{c}{AM1}&&\multicolumn{2}{c}{AM2}\\
\cline{3-3} \cline{5-6} \cline{8-9}
 &&$\alpha_0$ $\pm$ HW$_0$ && \hskip -10pt $\alpha_0$ $\pm$ HW$_0$ & \hskip -5pt $\alpha_1$ $\pm$ HW$_1$ && \hskip -10pt $\alpha_0$ $\pm$ HW$_0$ &  \hskip -5pt $\alpha_1$ $\pm$ HW$_1$\\ 
 &&(km s$^{-1}$ arcsec$^{-1}$) && \hskip -10pt (km s$^{-1}$ arcsec$^{-1}$) & \hskip -5pt (km s$^{-1}$) &&  \hskip -10pt (km s$^{-1}$ arcsec$^{-1}$) & \hskip -5pt (km s$^{-1}$ arcsec$^{-2}$) \\
 \tableline\\[-7.5 pt]
NGC 3031    & AW96a & (4.02 $\pm$ 0.51) $\times$ 10$^{-1}$ &&  \hskip -10pt (-1.20 $\pm$ 0.53) $\times$ 10$^{-1}$  & \hskip -1pt(1.50 $\pm$ \hskip 4pt 0.15) $\times$ 10$^{2}$ && \hskip -41pt 0.95 $\pm$ 0.09 & \hskip -0.5pt (-0.32  \hskip -0.3pt $\pm$ \hskip 4pt 0.54) $\times$ 10$^{-2}$ \\
NGC 3031    & AW96b & (2.12 $\pm$ 1.20) $\times$ 10$^{-1}$ &&  \hskip -10pt (-1.04 $\pm$ 0.64) $\times$ 10$^{-1}$  & \hskip -1pt(0.73 $\pm$ \hskip 4pt 0.13) $\times$ 10$^{2}$ &&\hskip -41pt 1.13 $\pm$ 0.24 & \hskip -0.5pt (-7.19 \hskip -0.3pt $\pm$ \hskip 4pt 1.84) $\times$ 10$^{-2}$ \\
NGC 3031 & H84 & (3.84 $\pm$ 0.60) $\times$ 10$^{-1}$ &&  \hskip -10pt (-4.57 $\pm$ 1.70) $\times$ 10$^{-1}$& \hskip -1pt (0.49 $\pm$ \hskip 4pt 0.10) $\times$ 10$^{2}$ &&\hskip -41pt 1.07 $\pm$ 0.16  &  \hskip -0.5pt  (-0.25   \hskip -0.3pt $\pm$ \hskip 4pt 0.06) $\times$ 10$^{-2}$ \\
NGC 3031 & YP99 & (3.11 $\pm$ 3.11) $\times$ 10$^{-1}$  &&  \hskip -10pt (-0.55 $\pm$ 4.20) $\times$ 10$^{-1}$ & \hskip -4.6pt (25.21 \hskip -0.0pt $\pm$ \hskip -0.4pt 26.49) $\times$ 10$^{2}$ &&\hskip -41pt 1.35 $\pm$ 0.84  &   \hskip -4.8pt (-14.7  \hskip 4pt $\pm$ 11.8)   \hskip 4.1pt $\times$ 10$^{-2}$  \\
NGC 2366 & $\mathcal{A}$ & (2.32 $\pm$ 0.73) $\times$ 10$^{-1}$ &&\hskip -41pt -1.48 $\pm$  0.82 & \hskip -26pt 3.00 $\pm$ \hskip 4pt 1.44 &&\hskip -41pt 1.71 $\pm$  0.72 &\hskip -0.5pt (-7.76 \hskip -0.3pt $\pm$ \hskip 4pt 3.76) $\times$ 10$^{-1}$\\
NGC 2366 & $\mathcal{B}$ & (2.66 $\pm$ 1.06) $\times$ 10$^{-1}$ &&\hskip -41pt -2.63 $\pm$ 2.96  & \hskip -26pt 3.93 $\pm$  \hskip 4pt 4.02 &&\hskip -41pt 2.51 $\pm$ 1.84  & \hskip -4.8pt (-15.3  \hskip 4pt  $\pm$ 12.5)  \hskip 4.1pt $\times$ 10$^{-1}$  \\
NGC 2366 & $\mathcal{C}$ & (2.13 $\pm$ 1.00) $\times$ 10$^{-1}$  &&\hskip -41pt -1.09 $\pm$ 0.69  & \hskip -26pt 3.60 $\pm$  \hskip 4pt 1.89 &&\hskip -41pt 1.21 $\pm$  0.53 & \hskip -0.5pt (-2.66 \hskip -0.3pt $\pm$ \hskip 4pt 1.42) $\times$ 10$^{-1}$  \\
NGC 2366 & $\mathcal{D}$ & (2.63 $\pm$ 1.24) $\times$ 10$^{-1}$  &&\hskip -41pt -0.21 $\pm$ 0.44  & \hskip -26pt 2.29 $\pm$  \hskip 4pt 2.09 &&\hskip -41pt 0.61 $\pm$ 0.35 &\hskip -0.5pt (-0.39  \hskip -0.3pt $\pm$ \hskip 4pt 0.38) $\times$ 10$^{-1}$  \\ 
DDO 154 &  & (1.76 $\pm$ 0.33) $\times$ 10$^{-1}$   &&  \hskip -10pt (-1.22 $\pm$ 0.29) $\times$ 10$^{-1}$ &\hskip -1pt (1.26 $\pm$ \hskip 4pt 0.46) $\times$ 10$^{1}$ &&\hskip -10pt  (2.15 $\pm$ 0.50) $\times$ 10$^{-1}$  & \hskip 0pt (-1.33 \hskip -0.5pt $\pm$ \hskip 4.3pt 1.36)  \hskip 0pt$\times$ 10$^{-6}$
\enddata
\end{deluxetable*} 

\begin{deluxetable*}{llccccccccc}[h!]
\tablecaption{Results for the $F$ Test} 
\tablewidth{0pt}
\startdata
\tableline\tableline\\[-7.5 pt]
Galaxy & Notes &Constant $\Omega_p$&&\multicolumn{3}{c}{AM1}&&\multicolumn{3}{c}{AM2}\\
\cline{3-3} \cline{5-7} \cline{9-11}
 && $\chi_\nu^{2}$ $\pm$ $\sigma$ && \hskip -6pt $\chi_\nu^{2}$  $\pm$ $\sigma$ & $F$ & $P_{\scriptsize{F}}$ && \hskip -8pt $\chi_\nu^{2}$  $\pm$ $\sigma$ & $F$ & $P_{\scriptsize{F}}$ \\ 
 &&&&&& (\%) &&&& (\%) \\
 \tableline\\[-7.5 pt]
NGC 3031    & AW96a & 25.8  \hskip 5.5pt $\pm$ 0.22 && 1.16 $\pm$ 0.22 &  \hskip -8.0pt 873  & $<$ 0.01 && 1.06 $\pm$ 0.22 & \hskip -3.5pt 955 & $<$ 0.01 \\
NGC 3031  & AW96b & \hskip -4.4pt 125  \hskip 12.3pt $\pm$ 0.47  && 4.28 $\pm$  0.50 &  \hskip -8.0pt 255 & $<$ 0.01 && 2.12 $\pm$ 0.50 & \hskip -3.5pt 523 & $<$ 0.01 \\
NGC 3031    & H84 & 24.0 \hskip 5.5pt $\pm$ 0.22 && 1.29 $\pm$ 0.22 &  \hskip -8.0pt 874 & $<$ 0.01 && 1.09 $\pm$ 0.22 & \hskip -7.5pt 1036 & $<$ 0.01  \\
NGC 3031    & YP99 & \hskip 0pt 73.0 \hskip 5.5pt $\pm$ 0.82 && 3.40 $\pm$ 1.00 &\hskip 2.9pt 52.8  & \hskip 9.5pt 1.84  && 2.19 $\pm$ 1.00 & \hskip 7.5pt 97.9 & \hskip 9.5pt 1.01  \\
NGC 2366    & $\mathcal{A}$ & \hskip 4.3pt 9.54 \hskip 1.3pt $\pm$ 0.28 && 0.90 $\pm$ 0.29 &  \hskip -8.0pt 241  & $<$ 0.01 && 0.87 $\pm$ 0.29 & \hskip -3.5pt 250 & $<$ 0.01 \\
NGC 2366   & $\mathcal{B}$  & \hskip 8.5pt 6.97 \hskip 1.4pt $\pm$ 0.39$^*$ && 0.65 $\pm$ 0.41 &  \hskip -8.0pt 127  & $<$ 0.01 && 0.83 $\pm$ 0.41& \hskip 7.5pt 97.6 & $<$ 0.01  \\
NGC 2366    & $\mathcal{C}$ & 13.8  \hskip 5.5pt $\pm$ 0.41 && 1.02 $\pm$ 0.43 &  \hskip -8.0pt 151  & $<$ 0.01 && 1.09 $\pm$ 0.43 & \hskip -3.5pt 141 & $<$ 0.01  \\
NGC 2366   & $\mathcal{D}$ & \hskip 4.3pt 5.55 \hskip 1.3pt $\pm$ 0.41 && \hskip 4.5pt 3.52 $\pm$ 0.43$^*$ &  \hskip 11.5pt 7.94  & \hskip 9.5pt 1.67 && 3.07 $\pm$ 0.43 & \hskip 7.5pt 10.7 & \hskip 9.5pt 0.75  \\
DDO 154    &  &  \hskip 4.3pt 0.83 \hskip 1.3pt $\pm$ 0.29 && 0.65 $\pm$ 0.30 &  \hskip 11.5pt 6.71 & \hskip 9.5pt 1.67 && \hskip 4.3pt 0.39 $\pm$ 0.30$^*$ & \hskip 7.5pt 26.2 &  $<$ 0.01 
\enddata
\end{deluxetable*} 

Plots of $\Omega_p$ calculated from the coefficients for NGC 3031 are shown in Figure 8.  Included in the plots is the material speed, $\Omega$, and possible locations for 2-arm Lindblad resonance.  The material speed is calculated from the rotation curve by de Blok et al. (2008).  The figure shows consistent plots among the four data sets for each of the different forms of $\Omega_p$.  The plots for AM1 show the same radial behavior as $\Omega$, with smaller values of $\Omega_p$ in the inner half of the disk for the AW96b data set, and outer half of the disk for the YP99 data set.  The solutions using $n_1$ more closely resemble the concave up behavior of $\Omega$ in the inner half of the disk whereas solutions using $n_2$ more closely resemble the linear behavior of $\Omega$ in the outer half of the disk.  The plots of AM1 and AM2 for the AW96a, AW96b, and H84 data sets do not show a unique point of corotation resonance.   For the same plots the existence of an outer 2-arm Lindblad resonance is inconclusive because the radius where $\Omega_p$ intersects possible locations for such a resonance also show corotation.  For both the AW96a and H84 data sets, the plots of AM1 and AM2 do not show an inner 2-arm Lindblad resonance.  The existence of an inner 2-arm Lindblad resonance is not ruled out by AM2 for the AW96b data set due to the slightly larger confidence bands.  The existence or absence of these resonances in plots of AM1 and AM2 are not as well constrained  for the YP99 data set due to the larger confidence bands.

\begin{figure*}[ht!]
\centering
\includegraphics[width=1\textwidth]{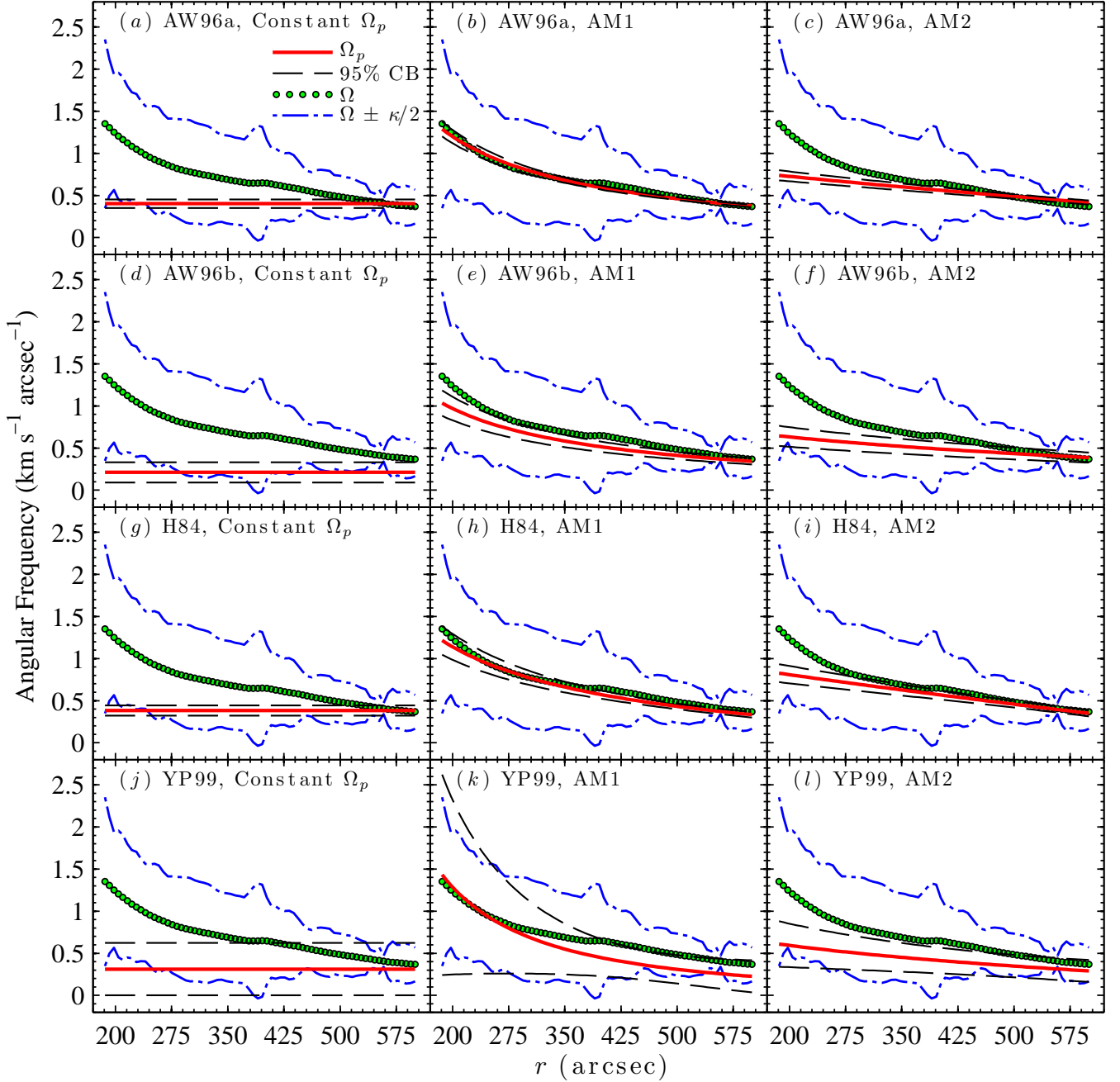} 
\caption{Plots of $\Omega_p$ and possible locations for resonance in NGC 3031.  The solid lines (colored red in the online version) and dashed lines show $\Omega_p$ and the 95\% confidence bands, respectively.  The open circles (filled in green in the online version) show the material speed, $\Omega$.  The dash-dot lines (colored blue in the online version) show possible locations of 2-arm Lindblad resonance.   From the top row to the bottom, shown are the results for the AW96a, AW96b, H84, and YP99 data sets.  The left, center, and right columns of panels show the results for a constant $\Omega_p$, AM1, and AM2, respectively.  Note how similar the plots for the AM1 and AM2 models are to $\Omega$.} 
\end{figure*}

For all four data sets, and both AM1 and AM2, the values of $P_{F}$ are small enough to rule out the null hypothesis that $\alpha_1$ = 0.  This is convincing evidence that the pattern is shearing.  Note that this conclusion is consistent with the results of the $KS$ test applied to the constant $\Omega_p$ model.  The results for the constant $\Omega_p$ models are therefore estimates of the mean of $\Omega_p$.  For each data set it is not clear whether AM1 or AM2 provides the better solution, and thus the better estimate of $\Omega_p$.  Their values of $\chi^{2}_\nu$ are quite similar, especially given their estimated 1$\sigma$ uncertainties.   Despite this ambiguity, it is possible to infer from the the above conclusions and the plots of AM1 and AM2 that there are no clear signs of 2-arm Lindblad resonances or a unique location of corotation resonance for the spiral pattern of NGC 3031.  It is also possible to infer that the speed of the spiral pattern more closely resembles the speed of the material than the speed of a rigidly rotating pattern.

Figure 9 shows how an incorrect PA could affect the conclusions for the NGC 3031 data sets.  In the figure (and similar figures for NGC 2366 and DDO 154) are values of $n$, $D$, $\chi_{\nu}^{2}$, $F$, and the mean of $\Omega_p$ that result from using a range of PA within $\pm$ 2$\sigma_{\mbox{\tiny PA}}$ of the assumed PA.  The deviations for the mean of $\Omega_p$ are shown because they are indicative of the amount of deviation in AM1 and AM2.  The values of $n$ and $\chi_{\nu}^{2}$ shown in the figure for AM1 and AM2 do not vary as much for the AW96a and H84 data sets as they do for the AW96b and YP99 data sets.  For all four data sets the values of $D$ are too small to rule out the null hypothesis that the normalized residuals for AM1 and AM2 have a standard normal distribution, except for the H84 data set at $\bigtriangleup$PA $\sim$ -1.5$\degr$.  This means that the AM1 and AM2 for the H84 data sets may not be the true model if the assumed PA is incorrect by $\sim$ 1.5$\degr$.  Determining whether the pattern is shearing, however, does not require that AM1 or AM2 are the true model.  A $\bigtriangleup$PA as large as $\pm$ 2$\sigma_{\mbox{\tiny PA}}$ would not change the results of the $F$ test for the AW96a, AW96b and H84 data sets.  Although some of the AM1 $F$ values around $\bigtriangleup$PA $\lesssim$ 1$\degr$ for AW96b appear close to the value at which $P_{F}$ = 5$\%$ in the figure, none of the $F$ values are actually small enough that $P_{F}$ $\geqslant$ 1$\%$.  There are 6 points for the YP99 data set where that the $F$ test is inconclusive.  These occur at $\bigtriangleup$PA = -0.94$\degr$ for AM1 and -0.73$\degr$ $\leqslant$ $\bigtriangleup$PA $\leqslant$ -0.31$\degr$ for AM2.  The only point in Figure 9 where that the $F$ test cannot rule out the null hypothesis that $\alpha_1$ = 0 occurs at $\bigtriangleup$PA = -0.83$\degr$ for the YP99 data set and AM1.

\begin{figure*}[ht!]
\centering
\hskip 100pt
\includegraphics[width=1\textwidth]{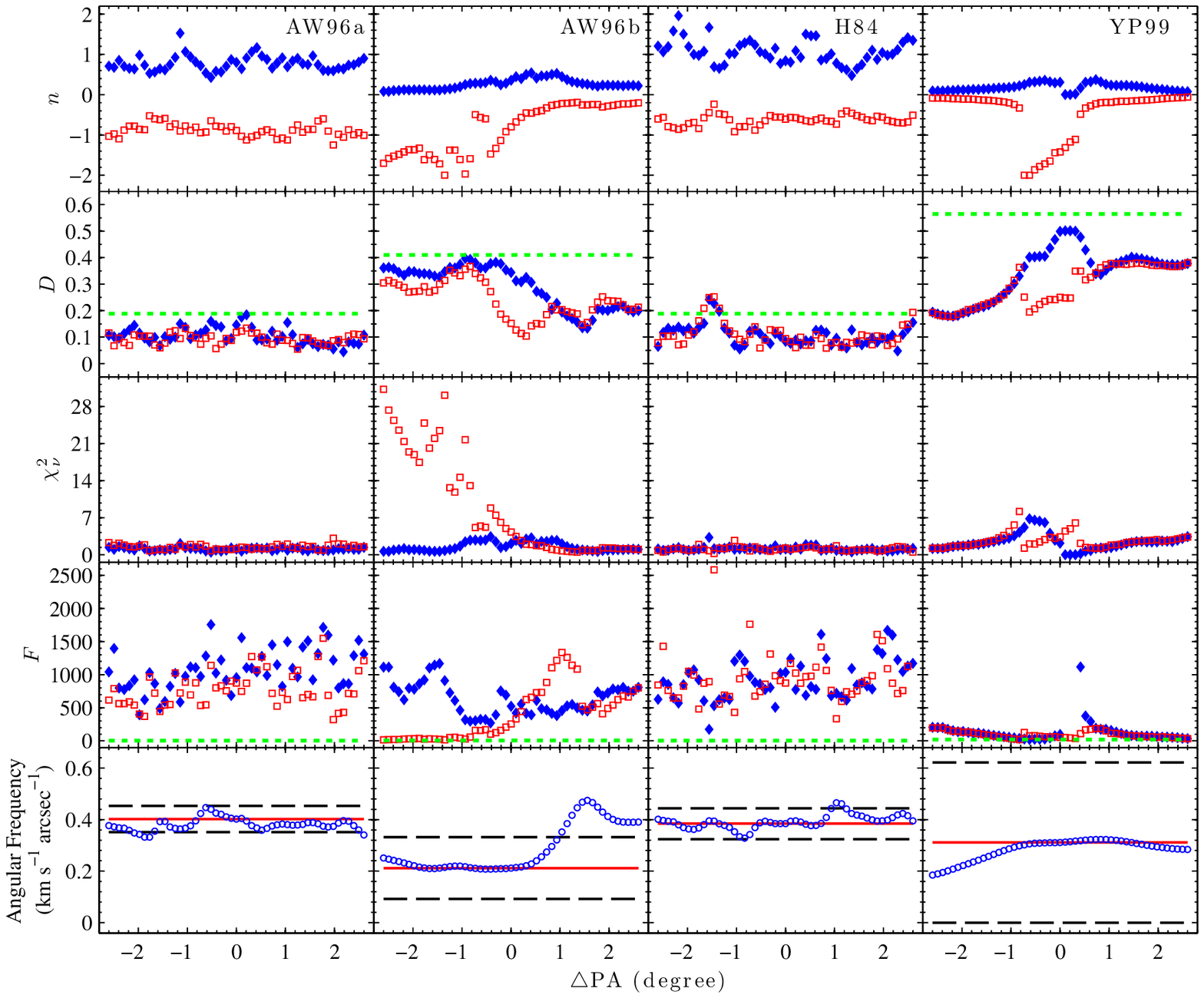} 
\caption{Plots showing how an incorrect PA could affect the findings for the NGC 3031 data sets.  From left to right are sets of plots for the AW96a, AW96b, H84, and YP99 data sets.  From top to bottom, each set of plots shows values of $n$, $D$, $\chi_{\nu}^{2}$, and $F$ where that the values for AM1 and AM2 are shown as open squares (colored red in the online version) and filled in diamonds (colored blue in the online version), respectively. The bottom plot in each set shows the mean of $\Omega_p$ as open circles (colored blue in the online version).  The dashed lines (colored green in the online version) included in the plots of $D$ and $F$ show the value where $P_{D}$ and $P_{F}$ are equal to 5\%.  Note that smaller values of $D$ correspond to larger values of $P_{D}$, and vice versa for $F$ and $P_{F}$.  The solid lines (colored red in the online version) and dashed lines included in the plots of the mean of $\Omega_p$ are the values obtained using the assumed PA and their 95\% confidence bands, respectively.}
\end{figure*}

Two points are worth emphasizing regarding how an incorrect PA could affect the conclusions for NGC 3031: 

\hskip 0pc \begin{minipage}[ht!]{0.1\textwidth} \vskip 0pt 
1.)  
\end{minipage} 

\hskip 1.4pc \begin{minipage}[h!]{0.427\textwidth} \vskip -10.5pt
All of the values of $F$ shown for three out of four data sets are large enough to conclusively rule out the null hypothesis that $\alpha_1$ = 0.  For all four data sets, 98.5$\%$ of the $F$ values are large enough to conclusively rule out the null hypothesis.   
\end{minipage}

\hskip 0pc \begin{minipage}[ht!]{0.1\textwidth} \vskip 0pt 
2.)  
\end{minipage} 

\hskip 1.4pc \begin{minipage}[h!]{0.427\textwidth} \vskip -10.5pt
Most of the mean $\Omega_p$ results (87.7$\%$) are within the 95$\%$ confidence bands for the mean of $\Omega_p$ obtained using the adopted PA.  The most deviant points occur for the AW96b data set, and these are larger than this data set's mean value for $\Omega_p$.  Such an increase in the $\Omega_p$ results for the AW96b data set would only bring the results for this data set into better agreement with the results for the other data sets.
\end{minipage}

\noindent Therefore, the affect of an incorrect PA on the conclusions for NGC 3031 are insignificant.  The affect of an incorrect PA on the conclusions for NGC 1365 were also found to be insignificant in Paper I.  The robustness of the conclusions to an incorrect PA, in both cases, is attributable to the small uncertainties in the measured values of the PA, and the high quality of the H{\hskip 1.5pt \footnotesize  I} data.

\subsection{Results for NGC 2366}

Unlike NGC 3031, different regions of NGC 2366 produce different results.  Results are presented for four different regions: $|y|$ $\leqslant$ 400$\arcsec$, $|y|$ $\leqslant$ 200$\arcsec$, $-$400$\arcsec$ $\leqslant$ $y$ $\leqslant$ 0$\arcsec$, and 0$\arcsec$ $\leqslant$ $y$ $\leqslant$ 400$\arcsec$.  These are hereafter referred to as regions $\mathcal{A}$, $\mathcal{B}$, $\mathcal{C}$, and $\mathcal{D}$, respectively.  They provide 26, 14, 13, and 13 independent calculations of Equation (1), respectively.

In Figure 7 there are prominent peaks in the values of $P_{KS}$ for regions $\mathcal{A}$, $\mathcal{B}$, and $\mathcal{C}$.  The peaks for region $\mathcal{D}$ are not as prominent.  Similar to the plots for NGC 3031, the plots for all four regions of NGC 2366 show small $P_{KS}$ near $n$ =  0.  For regions $\mathcal{A}$, $\mathcal{B}$, and $\mathcal{C}$, the values of $P_{KS}$ for the constant $\Omega_p$ model are small enough to rule out the null hypothesis that the normalized residuals have a standard normal distribution.  This is convincing evidence that the constant $\Omega_p$ model is not the true model for these regions.  For region $\mathcal{D}$, $P_{KS}$ for the constant $\Omega_p$ model is too large to rule out the null hypothesis that the normalized residuals have a standard normal distribution.  Nothing can be inferred from the $KS$ test about the constant $\Omega_p$ model in region $\mathcal{D}$.  

Plots of $\Omega_p$ calculated from the coefficients for NGC 2366 are shown in Figure 10.  The material speed in the figure is calculated from the rotation curve by Oh et al. (2008; hereafter O08).  Possible locations for 2-arm lindblad resonances are included in the plots (and also for DDO 154) to create the same sense of scale as the plots for NGC 3031.  All of the regions show consistent plots for the constant $\Omega_p$ model.  All of the regions also show corotation in the inner half of the disk for AM1 and AM2.  Note that for each region, their plots for AM1 and AM2 are similar although their values of $n$ have opposite sign.  There are clear differences among the different regions for the plots in the outer half of the disk.  These are the most prominent for regions $\mathcal{C}$ and $\mathcal{D}$.  In region $\mathcal{C}$ the plots for AM1 and AM2 approach $\Omega_p$ = 0 at $\sim$ 400$^{\arcsec}$, which is $\sim$ 0.16 km s$^{-1}$ arcsec$^{-1}$ less than $\Omega$.  In region $\mathcal{D}$ the results for AM1 and AM2 are similar to $\Omega$ for the entire range in $r$ shown. 

\begin{figure*}[ht!]
\centering
\includegraphics[width=1\textwidth]{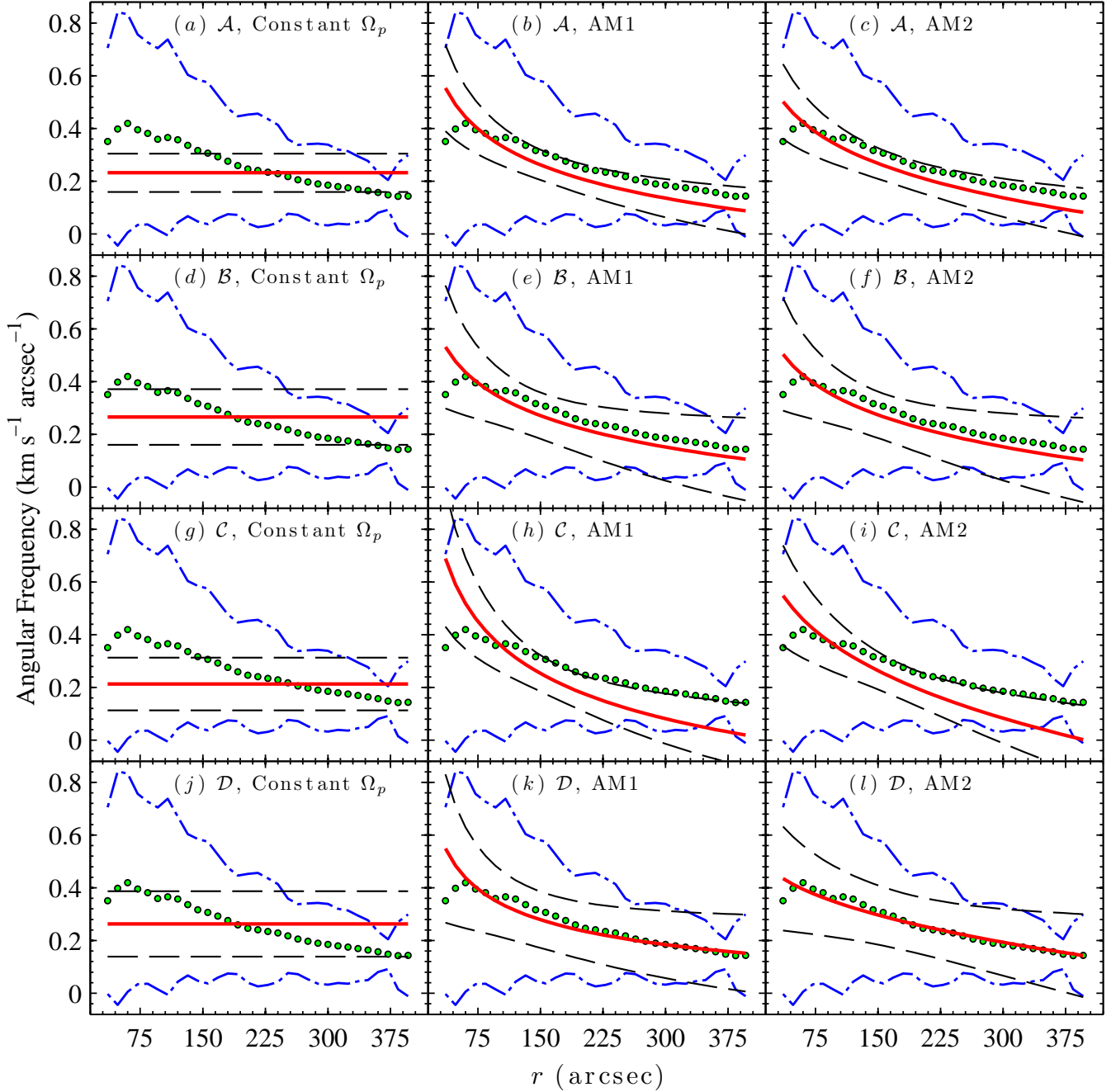} 
\caption{Plots of $\Omega_p$ and possible locations for resonance in NGC 2366.  The figure is formatted in the same way as Figure 8.  From the top row to the bottom, shown are the results for regions $\mathcal{A}$, $\mathcal{B}$, $\mathcal{C}$, and $\mathcal{D}$.  The left, center, and right columns of panels show the results for a constant $\Omega_p$, AM1, and AM2, respectively. Note the differences for the results shown in panels ($h$) and ($k$), and likewise for panels ($i$) and ($l$).}
\end{figure*}

For all four regions, and both AM1 and AM2, the values of $P_{F}$ are small enough to rule out the null hypothesis that $\alpha_1$ = 0.  For regions $\mathcal{A}$, $\mathcal{B}$, and $\mathcal{C}$, the results of the $F$ test are consistent with the results of the $KS$ test applied to the constant $\Omega_p$ model.  The results of the $F$ test for region $\mathcal{B}$, however, are unreliable because the distribution of the normalized residuals for the constant $\Omega_p$ model failed a Lilliefors test for normality ($P_L$ =1.79\%).  The $F$ test result for AM1 in region $\mathcal{D}$ may also be unreliable because the results of a Lllliefors test applied to its normalized residuals is inconclusive ($P_L$ = 5.35\%).  From the results for the $F$ test,  the results of the $KS$ test for the constant $\Omega_p$ model, and the plots of $\Omega_p$, it is possible to infer that the clumps of H{\hskip 1.5pt \footnotesize  I} in NGC 2366 are corotating with the material in the inner half of the disk, and that there are clumps of H{\hskip 1.5pt \footnotesize  I} in the outer half of the disk in region $\mathcal{C}$ that are rotating more slowly than the rest of the material. 

\begin{figure*}[ht!]
\centering
\includegraphics[width=1\textwidth]{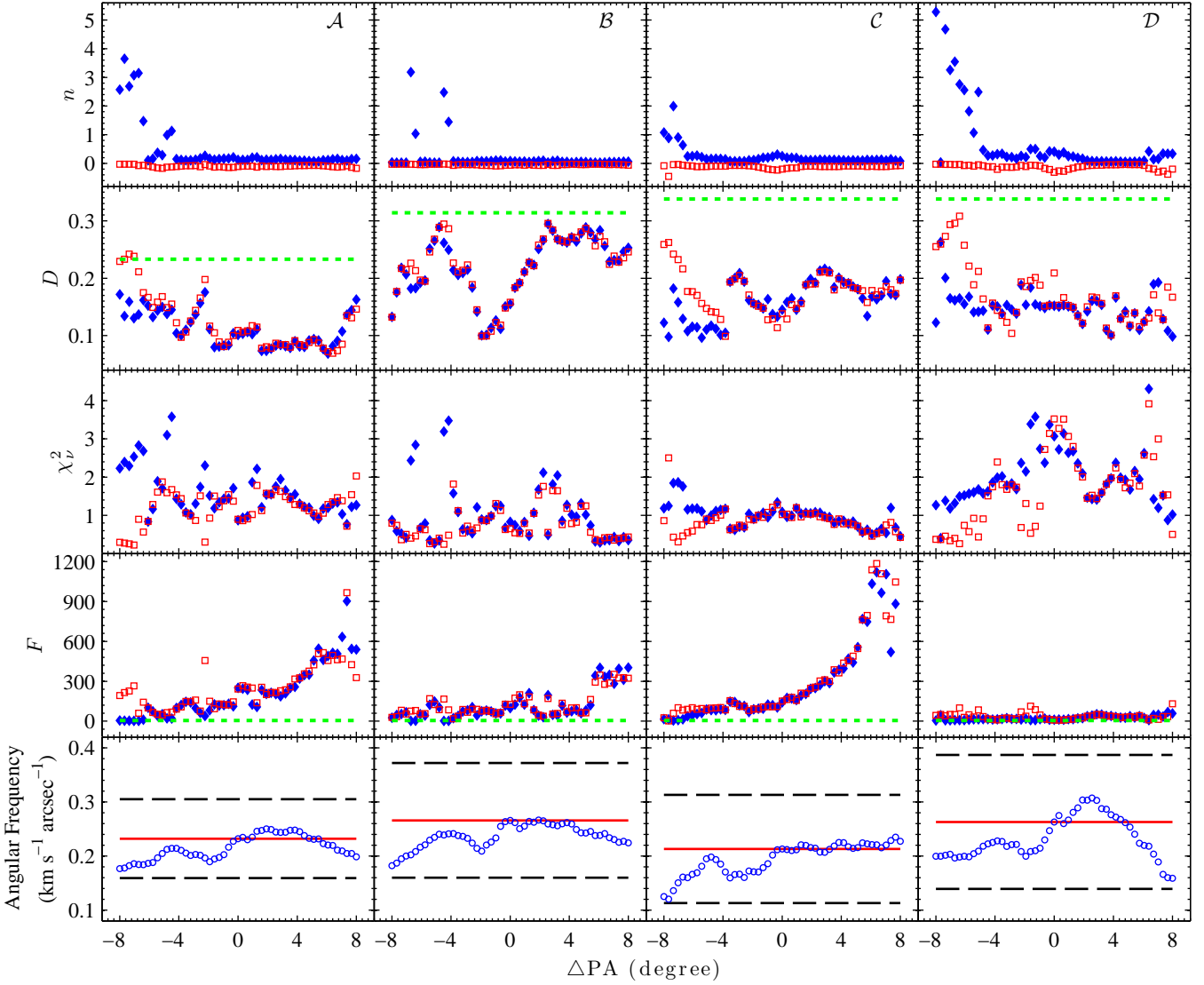} 
\caption{Plots showing how an incorrect PA could affect the findings for the NGC 2366 data sets.  The figure is formatted in the same way as Figure 9. From left to right are sets of plots for regions $\mathcal{A}$, $\mathcal{B}$, $\mathcal{C}$, and $\mathcal{D}$, respectively.} 
\end{figure*}

The results for NGC 2366 show that the solution method can detect different values of $\Omega_p$ in different regions of a galaxy.  This demonstrates the usefulness of the solution method for determining if there is a single, global pattern speed.  The difference for regions $\mathcal{C}$ and $\mathcal{D}$ may be attributed to the non-circular motions in the velocity field that are present in region $\mathcal{C}$.  These are easily identified in Figure 2 by the kinks in the contours of constant $v_y$ around $y$ $\sim$ 200$^{\arcsec}$ - 400$^{\arcsec}$.  The non-circular motions were removed from the velocity field when O08 derived the rotation curve used in this paper for calculating $\Omega$.  The difference between the rotation curve of O08, and one obtained without the non-circular motions removed (see Figure 13 of O08), is not large enough to account for the difference shown for $\Omega$ and $\Omega_p$ in panels (h) and (i) of Figure 10.  The solution method is clearly detecting something other than $\Omega_p$ $\approx$ $\Omega$.

Figure 11 shows how an incorrect PA could affect the conclusions for the different regions of NGC 2366.  For all four regions, the values of $n$ do not change significantly, except for AM1 at $\bigtriangleup$PA $\lesssim$ -4$\degr$.  Almost all of the values of $D$ that are shown are too small to rule out the null hypothesis that the normalized residuals for AM1 and AM2 have a standard normal distribution, consistent with the values for these models in Table 3.  The values of $\chi_{\nu}^{2}$ for regions $\mathcal{A}$, $\mathcal{B}$, and $\mathcal{C}$, are fairly consistent, especially given their uncertainties.  The values of $F$ are all quite large for $\mathcal{A}$ and $\mathcal{C}$.  The smallest values of $F$ occur for region $\mathcal{D}$.  Note that for all four regions, the values shown for the mean of $\Omega_p$ are within the 95\% confidence bands for the value obtained using the adopted PA. It is reasonable to conclude that the accuracy of the adopted PA is not likely to significantly affect any of the conclusions for NGC 2366. 

\subsection{Results for DDO 154}

Similar to NGC 3031, different regions: $|y|$ $\leqslant$ 400$\arcsec$, $|y|$ $\leqslant$ 200$\arcsec$, $-$400$\arcsec$ $\leqslant$ $y$ $\leqslant$ 0$\arcsec$, and 0$\arcsec$ $\leqslant$ $y$ $\leqslant$ 400$\arcsec$, produce consistent results.  Therefore, only the results for most of the disk, $|y|$ $\leqslant$ 400$\arcsec$, are presented.  This region provides 24 independent calculations of Equation (1).

\begin{figure*}[ht!]
\centering
\includegraphics[width=1\textwidth]{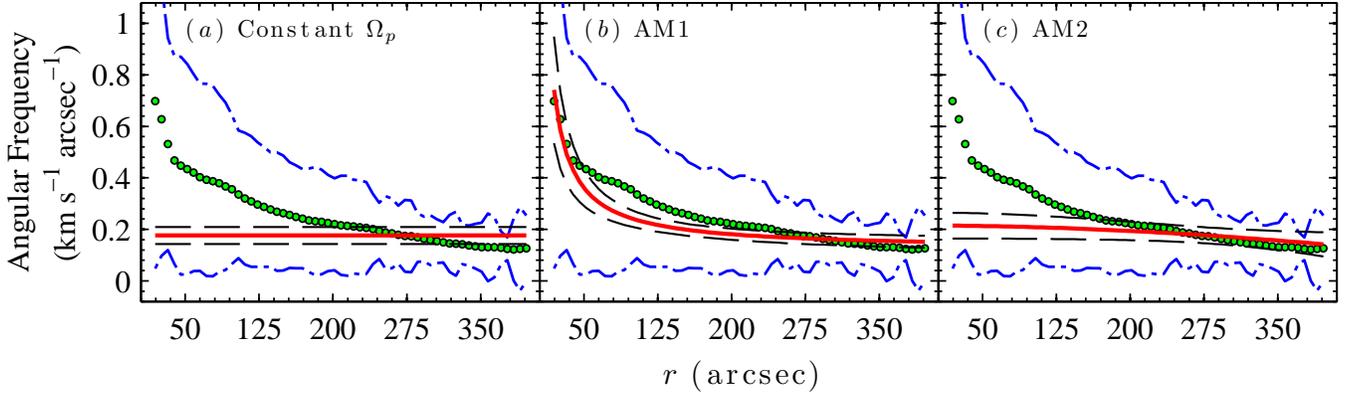} 
\caption{Plots of $\Omega_p$ and possible locations for resonance in DDO154.  The figure is formatted in the same way as Figure 8.} 
\end{figure*}

The peaks in the distribution of $P_{KS}$ shown in Figure 7 are not as narrow as they are in most of the other plots.  Similar to most of the other plots, however, $P_{KS}$ is small near $n$ = 0.  Note that the value of $P_{KS}$ for the constant $\Omega_p$ model is the largest of the values of $P_{KS}$ shown in Table 3 for the constant $\Omega_p$ model.  It is too large to rule out the null hypothesis that the distribution of the normalized residuals is a standard normal distribution for the constant $\Omega_p$ model.  

Plots of $\Omega_p$ calculated from the coefficients are shown in Figure 12.  The material speed in the figure is calculated from the rotation curve by  Oh et al. (2011).  All three models of $\Omega_p$ are quite similar in the outer half of the disk.  The plots for the constant $\Omega_p$ model and AM2 are consistent with each other to within their 95\% confidence bands.  The plot for AM1 differs from the other two by showing corotation in the central part of the galaxy.  None of the plots show corotation at $\sim$ 100$^{\prime\prime}$.  

\begin{figure}[t!]
\centering
\includegraphics[width=0.48\textwidth]{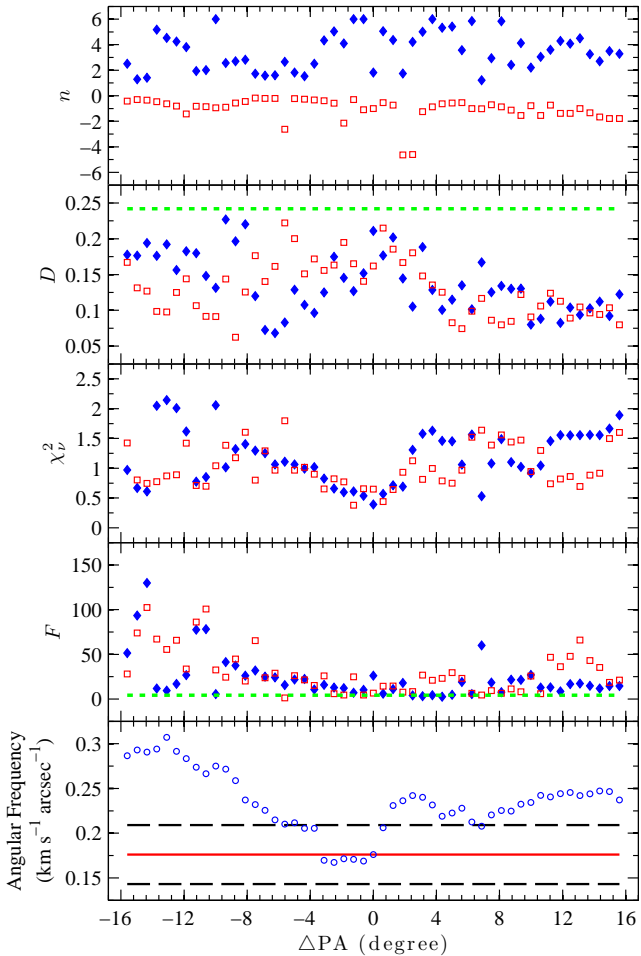} 
\caption{Plots showing how an incorrect PA could affect the findings for DDO 154.  The figure is formatted in the same way as Figure 9. } 
\end{figure}

The values of $P_{F}$ in Table 5 for AM1 and AM2 are small enough to rule out the null hypothesis that $\alpha_1$ = 0.  This result for AM2 may be unreliable because the results of a Lllliefors test applied to its normalized residuals is inconclusive ($P_L$ = 4.62\%).  Also note that the solution for this model has a value of $\chi_{\nu}^{2}$ that is smaller than 1 by more than 2$\sigma$.    From the results for the $F$ test, the results of the $KS$ test for the constant $\Omega_p$ model, and the plots of $\Omega_p$, the clumps of H{\hskip 1.5pt \footnotesize  I} in DDO 154 are corotating in the central and outer parts of the disk.  Some of the clumps of H{\hskip 1.5pt \footnotesize  I} at  $\sim$ 100$^{\prime\prime}$ are rotating more slowly than the material.  The result for the constant $\Omega_p$ model is the mean of $\Omega_p$.

Figure 13 shows how an incorrect PA could affect the conclusions for DDO 154.  The results for DDO 154 are much more dependent on the value of the adopted PA than are the results for NGC 3031 or NGC 2366.  The values of $n_2$ vary by as much as a factor of 6.  The values of $\chi_{\nu}^{2}$ vary by almost as much, with the smallest values occurring around $\bigtriangleup$PA $\sim$ 0$\degr$.  There are values of $\bigtriangleup$PA where that the results of the $F$ test are inconclusive, or that $F$ is too small to rule out the null hypothesis that $\alpha_1$ = 0.  The results for the constant $\Omega_p$ model deviates quite significantly beyond the 95\% confidence intervals for the value obtained using the adopted PA.  It must be concluded that the accuracy of the adopted PA affects some of the conclusions for DDO 154, particularly whether the solution using a constant $\Omega_p$ model, or one of the alternative models, provides a better solution for Equation (1).  

\section{DISCUSSION}

\subsection{How Different Properties of a Data Set Affect the Results}

The previous section shows the results from applying the solution method to data sets with different angular resolutions, spectral resolutions, and sensitivities to structures on different scales.   Results are also shown for galaxies without coherent or well-organized patterns.  Knowing how these properties of a data set affect the results is important for choosing data in future applications of the solution method.

Of all the differences in the data sets, the angular resolution had the most noticeable affect on the results by setting the number of independent calculations of Equation (1). The most extreme cases can be seen in Figure 8 for the NGC 3031 data sets.  The results for the YP99 data set demonstrate that a constant $\Omega_p$, or the mean if there is evidence for a radial dependence, can be found with only 4 independent calculations of Equation (1), but more calculations would be desirable in order to reduce the uncertainty.  In order to determine whether there is a single, global pattern speed (i.e., $\Omega_p$ is the same in different regions of a galaxy), a number of independent calculations similar to what was used for AW96b and NGC 2366 should be sufficient, given similar uncertainties.  

There is no noticeable affect on the results due to spectral resolution.  Consistent results were obtained for the AW96a and H84 data sets despite a factor of $\sim$ 4 difference in the channel separation.  Also consider that results with small uncertainties were obtained in Paper I for NGC 1365 using maps made from a 20.84 km s$^{-1}$ channel separation.  The large channel separation for the NGC 1365 data may have been compensated for by having 62 independent calculations of Equation (1) for the whole disk.

The results for the three different data sets of NGC 3031 are remarkably consistent despite their different sensitivities to structures on different scales.  The results for the AW96a and H84 data sets are almost identical even though they are more sensitive to small and large scales, respectively, and the AW96a data set is missing some flux.  The results for the data sets of NGC 3031 with the largest angular resolution, AW96b and YP99, are slightly less than the results for AW96a and H84.  The smaller number of independent calculations of Equation (1) for the AW96b and YP99 data sets, and consequently larger uncertainties, may be the cause for this small difference.   

The results for NGC 2366 and DDO 154 demonstrate that the solution method can produce meaningful results that are simple to interpretin the absence of a coherent or well-organized pattern.  This shows that the solution method is not limited to grand-design patterns.  It could be used to find $\Omega_p$ for multiple armed and flocculent spiral galaxies as well.  For these types of spiral galaxies the solution method would be measuring $\Omega_p$ for different arms, or clumps of higher density material.  Although this may violate the assumption of a well-defined pattern speed by TW84 (i.e., the surface brightness is assumed to be constant in a frame rotating at an angular speed $\Omega_p$), the TW84 equations are generalizable to a frame that is rotating at a variable angular speed $\Omega_p$($r$) (Engstr\"{o}m 1994).  Whether the results for galaxies without a coherent or well-organized pattern have any physical meaning can be determined using a variety of statistical tools such as $\chi^{2}_\nu$ and the Lilliefor's test used in this paper.  A very large or small value of $\chi^{2}_\nu$, or a small enough value of $P_{L}$ for the normalized residuals, could mean that Equation (1) is a poor description of the dynamics of a galaxy, or part of a galaxy, for example.   

This paper shows that having a sufficient number of independent calculations is important for applying the solution method.  This limits the use of H{\hskip 1.5pt \footnotesize  I} radio synthesis data to nearby galaxies.  For example, the resolution of the C-array of the VLA for H{\hskip 1.5pt \footnotesize  I} observing, which has a typical FWHM of $\sim$ 13$\arcsec$ for the synthesized beam, would be adequate for many nearby spiral galaxies.  Radio synthesis data for tracers of molecular gas, which have a higher frequency, and thus a smaller synthesized beam (typically as small as sub arcsecond to a few arcseconds at $\sim$ 100 GHz), would be useful for applying the method to more distant galaxies that have a smaller angular size.  The resolution of optical data, such as Fabry-Perot H$\alpha$ data, is typically $\sim$ 1$\arcsec$ - 2$\arcsec$ due to atmospheric seeing.  Such data would also expand the applicability of the solution method to galaxies with a smaller angular size.

\subsection{A Closer Look at the Findings for NGC 3031}

There is convincing evidence in the form of the $F$ test that the spiral pattern in NGC 3031 is shearing.  In fact, the values of $F$ are quite large.  This confirms the discovery of shear in the pattern by Westpfahl (1998).  The pattern speed is very similar to the material speed, and there are no clear indications of unique corotation or Lindblad resonances.  The same conclusions were found for the spiral pattern of NGC 1365.  Note that the results of the $F$ test are complimented by the results of the $KS$ test for the constant $\Omega_p$ model, from which it can be inferred  that the constant $\Omega_p$ model is not the true model.  It is important to point out that these findings for both NGC 3031 and NGC 1365 are based on calculations that are independent of the assumptions made in spiral structure theories.  

The results for NGC 3031 are similar to the sheared gravitational instabilities of Goldreich \& Lynden-Bell (1965) and the swing amplifier of Toomre (1981).  Close passage of the companion galaxies NGC 3034 and NGC 3077 could have caused the swing amplification in NGC 3031.  Several models agree that these galaxies passed to within 20 kpc of NGC 3031 about 600 Myr ago (Van der Hulst 1977; Cottrell 1976, 1977; Gottesman \& Weliachew 1975, 1977; and Brouillet et al. 1991).  Simulations show that such interactions can produce grand design patterns that shear over time (S. Oh 2008; Dobbs et al. 2010; Struck et al. 2011).  This implies that the spiral pattern in NGC 3031 is a transient feature that is driven by interactions with its companions, and in this way the winding dilemma is avoided.  

The results in this paper and in Paper I are part of a growing body of evidence that some grand-design spiral patterns are shearing, transient features.  Some of this evidence is coming from simulations, which have yet to produce persisting patterns (e.g., Berman et al. 1978;  Sellwood \& Carlberg 1984; Sorensen 1985; Bottema 2003; Dobbs \& Bonnell 2008; S. Oh et al. 2008; Dobbs et al. 2010; Struck et al. 2011; Wada et al. 2011; Sellwood 2011 for a review).  Of the more recent simulations, Wada et al. (2011) showed that the shearing, short lived ($\sim$ 100 Myr), recurring spiral arms in their simulations of a live disk have a pattern speed that is similar to the material speed. 

Simulations are useful tools for forming hypothesis about spiral structure, but a true understanding must ultimately come from observations.  In Paper I it was pointed out that the original solution method of TW84 produced results for at least 7 nearby spiral galaxies that are consistent with a shearing pattern.  Foyle et al. (2011) also found evidence against long-lived, rigidly rotating patterns in spiral galaxies.  They showed that the angular offsets between star formation tracers in 11 nearby spiral galaxies, including NGC 3031, are not systemically ordered as predicted from a long-lived, rigidly rotating pattern. 

It is difficult to reconcile the findings in this paper for NGC 3031 with the density wave theory.  All versions of the theory use as a foundation the premise that the pattern is rotating approximately rigidly (Bertin \& Lin 1996, Chapter 2; Bertin 2000, Chapter 15; Binney \& Tremaine 2008, Chapter 6).  One possibility is that $\Omega_p$ is a step function, as advocated for by Meidt et al. (2008) and Meidt et al. (2009).  The solution method used in this paper would not be able to detect such step functions because continuity is assumed for AM1 and AM2.  The forms used in Paper I, however, are capable of detecting such step functions, and did not find evidence for this form of $\Omega_p$ in the spiral pattern of NGC 1365.  The same forms of $\Omega_p$ used in Paper I were also tried for the NGC 3031 data sets, and there is no sign of a step function.  Most importantly, note that there are no discontinuities in the spiral pattern of NGC 3031 as one would expect if $\Omega_p$ is indeed a step function.  

\section{SUMMARY}

In this paper the method derived in Paper I for solving the pattern speed equations of TW84 is applied to NGC 3031, NGC 2366, and DDO 154 in order to show how different properties of a data set affect the results.  The conclusions are as follows:

\hskip 0pc \begin{minipage}[ht!]{0.1\textwidth} \vskip 0pt 
1.)  
\end{minipage} 

\hskip 1.4pc \begin{minipage}[h!]{0.427\textwidth} \vskip -10.5pt
Four different data sets of NGC 3031 produce consistent results despite differences in angular resolution, spectral resolution, and sensitivities to structures on different scales.  They all produce convincing evidence for shear in the pattern.  The pattern speed is more similar to the material speed than it is to the speed of a rigidly rotating pattern.  There are no unique points of corotation or 2-arm Lindblad resonances.  
\end{minipage}

\hskip 0pc \begin{minipage}[ht!]{0.1\textwidth} \vskip 0pt 
2.)  
\end{minipage} 

\hskip 1.4pc \begin{minipage}[h!]{0.427\textwidth} \vskip -10.5pt The angular resolution had the most noticeable affect on the results by setting the number of independent calculations of Equation (1).  This is important to consider when selecting data for future \linebreak
\end{minipage} 
\\ 

\hskip 1.4pc \begin{minipage}[h!]{0.427\textwidth} \vskip 0.1pt applications of the solution method, especially for determining the radial behavior of $\Omega_p$, and whether there is a single, global pattern speed.  Lower angular resolution data sets give few independent rows of data, so are less useful for determining $\Omega_p$ as a function of radius.
\end{minipage} 

\hskip 0pc \begin{minipage}[ht!]{0.1\textwidth} \vskip -0pt 
3.)  
\end{minipage} 

\hskip 1.4pc \begin{minipage}[h!]{0.427\textwidth} \vskip -10.5pt
The results for NGC 2366 show that the solution method can detect different pattern speeds in different parts of a galaxy.
\end{minipage} 

\hskip 0pc \begin{minipage}[ht!]{0.1\textwidth} \vskip -6pt 
4.)  
\end{minipage} 

\hskip 1.4pc \begin{minipage}[h!]{0.427\textwidth} \vskip -10.5pt
The results for NGC 2366 and DDO 154 show that meaningful results that are simple to interpret can be obtained even in the absence of a coherent or well organized pattern.  The results for these galaxies also demonstrate that the solution method is not merely measuring the material speed, and thus serve as a counterexample to the results for NGC 3031.  
\end{minipage}

\acknowledgements

The authors acknowledge the helpful comments of the referee that improved this paper.  The authors also acknowledge Erwin de Blok and Se-Heon Oh for providing the rotation curve data and the uncertainties of the orientation parameters for NGC 3031, NGC 2366, and DDO 154.  Renee Andrae, Tim Shulze-Hartung, and Peter Melchoir provided some of the inspiration for the procedure used in this paper for finding models that are an alternative to a constant $\Omega_p$.  This research made use of the National Radio Astronomy Observatory which is a facility of the National Science Foundation operated under cooperative agreement by Associated Universities, Inc; and the NASA/IPAC Extragalactic Database which is operated by the Jet Propulsion Laboratory, California Institute of Technology, under contract with the National Aeronautics and Space Administration.

\end{document}